\documentclass[conference]{IEEEtran}
\IEEEoverridecommandlockouts
\AtBeginDocument{}
\renewcommand{\footnotesize}{\scriptsize}
\usepackage{lineno}
\usepackage{footnote}
\usepackage{commath}
\usepackage{alphalph}
\usepackage{enumitem}
\usepackage[T1]{fontenc}
\usepackage{subcaption}
\usepackage{dirtytalk}
\usepackage[labelformat=parens,labelsep=quad, skip=3pt]{caption}
\usepackage{color,soul}
\usepackage{cite}
\usepackage{amsmath,amssymb,amsfonts}
\usepackage{algorithmic}
\usepackage{textcomp}
\usepackage{xcolor}
\usepackage{comment}
\usepackage{graphicx}
\usepackage{tabularx,booktabs}
\usepackage{array}
\usepackage{bbding}
\usepackage{pifont}
\usepackage{wasysym}
\usepackage{amssymb}
\usepackage{amsmath} % Star
\usepackage{tabularx}
\usepackage{pgfplots}
\pgfplotsset{width=7cm,compat=1.8,tick label style={font=\small}}
\usepackage{pgfplotstable}
\usepackage{sfmath}
\usepackage{array, booktabs, makecell}
\usepackage{color}
\usepackage{csquotes}
\usepackage{url}
\usepackage{comment}
\usepackage{tikz}
\usepackage{balance}
\usepackage{listings}
\usepackage{booktabs} % used for \toprule in tables
\usepackage{cite}
\usepackage{soul} % highlighting
\usetikzlibrary{fit} %% Used for putting dotted box in image
\usetikzlibrary{positioning}
\usetikzlibrary{arrows}
\usetikzlibrary{shapes.multipart}
\usepackage{caption}
\usepackage{graphicx}
\usepackage{adjustbox}
\usepackage{float}
\usepackage{rotating}
\usepackage{tablefootnote}
\usepackage{nth}
\DeclareCaptionType{TextBox}

\usepackage[utf8]{inputenc}
\usepackage{dirtytalk}
\usepackage{tcolorbox}

\usepackage[british]{babel}
\usepackage{enumitem}

\newlist{SubItemList}{itemize}{1}
\setlist[SubItemList]{label={$-$}}

\let\OldItem\item
\newcommand{\SubItemStart}[1]{%
    \let\item\SubItemEnd
    \begin{SubItemList}[resume]%
        \OldItem #1%
}
\newcommand{\SubItemMiddle}[1]{%
    \OldItem #1%
}
\newcommand{\SubItemEnd}[1]{%
    \end{SubItemList}%
    \let\item\OldItem
    \item #1%
}
\newcommand*{\SubItem}[1]{%
    \let\SubItem\SubItemMiddle%
    \SubItemStart{#1}%
}%
\usepackage[bookmarks=false]{hyperref}
\usetikzlibrary{shapes.geometric, arrows} 

\tikzstyle{startstop} = [rectangle, rounded corners, minimum width=3cm, minimum height=1cm,text centered, draw=black, fill=black!10]

\tikzstyle{io} = [trapezium, trapezium left angle=70, trapezium right angle=110, minimum width=3cm, minimum height=1cm, text centered,text width=2cm, draw=black]

\tikzstyle{process} = [rectangle, minimum width=3cm, minimum height=1cm, text centered, text width=3cm, draw=black]

\tikzstyle{decision} = [diamond, minimum width=3cm, minimum height=1cm, text centered,text width=2cm, draw=black]

\tikzstyle{arrow} = [thick,->,>=stealth]

\tikzstyle{database} = [cylinder, shape border rotate=90, draw=black,minimum height=2cm,minimum width=3cm, text centered,text width=0.6cm]
\usepackage{multirow} 
\usepackage{multicol}
\usepackage{varwidth} % Double column tables
\newcolumntype{L}[1]{>{\raggedright\arraybackslash}m{#1}} % raggedright= align left
\definecolor{Gray}{gray}{0.80} % the lower the #, the darker it gets
\newlength\q %% Used for many tables
\def\BibTeX{{\rm B\kern-.05em{\sc i\kern-.025em b}\kern-.08em
    T\kern-.1667em\lower.7ex\hbox{E}\kern-.125emX}}
    
        {\begin{itemize}\setlength{\itemsep}{0pt}\setlength{\parsep}{0pt}
	\setlength{\topsep}{0pt}\setlength{\parskip}{0pt}}
	{\end{itemize}}
	{\begin{enumerate}\setlength{\itemsep}{0pt}\setlength{\parsep}{0pt}
	\setlength{\topsep}{0pt}\setlength{\parskip}{0pt}}
	{\end{enumerate}}
	{\begin{enumerate}\setlength{\itemsep}{0pt}\setlength{\parsep}{0pt}
	\setlength{\topsep}{0pt}\setlength{\parskip}{0pt}\setlength{\leftmargin}{0pt}}
	{\end{enumerate}}   
\begin{document}

\title{ \huge Do Design Metrics Capture Developers Perception of Quality? An Empirical Study on Self-Affirmed Refactoring Activities\\
%\title{ \huge On the Impact of Refactoring on the Relationship between Quality Attributes and Design Metrics\\
}

\author{
\IEEEauthorblockN{Eman Abdullah AlOmar\IEEEauthorrefmark{1}, 
Mohamed Wiem Mkaouer\IEEEauthorrefmark{1},
Ali Ouni\IEEEauthorrefmark{2}, Marouane Kessentini\IEEEauthorrefmark{3}}
\IEEEauthorblockA{\IEEEauthorrefmark{1}Rochester Institute of Technology, NY, USA\\
\IEEEauthorrefmark{2}ETS Montreal, University of Quebec, Montreal, QC, Canada\\
\IEEEauthorrefmark{3} University of Michigan, Michigan, USA\\
\text{eman.alomar@mail.rit.edu, mwmvse@rit.edu}, \text{ali.ouni@etsmtl.ca},
\text{marouane@umich.edu}\\
}}

\begin{comment}
\author{\IEEEauthorblockN{1\textsuperscript{st} Given Name Surname}
\IEEEauthorblockA{\textit{dept. name of organization (of Aff.)} \\
\textit{name of organization (of Aff.)}\\
City, Country \\
email address}
\and
\IEEEauthorblockN{2\textsuperscript{nd} Given Name Surname}
\IEEEauthorblockA{\textit{dept. name of organization (of Aff.)} \\
\textit{name of organization (of Aff.)}\\
City, Country \\
email address}
\and
\IEEEauthorblockN{3\textsuperscript{rd} Given Name Surname}
\IEEEauthorblockA{\textit{dept. name of organization (of Aff.)} \\
\textit{name of organization (of Aff.)}\\
City, Country \\
email address}
\and
\IEEEauthorblockN{4\textsuperscript{th} Given Name Surname}
\IEEEauthorblockA{\textit{dept. name of organization (of Aff.)} \\
\textit{name of organization (of Aff.)}\\
City, Country \\
email address}
\and
\IEEEauthorblockN{5\textsuperscript{th} Given Name Surname}
\IEEEauthorblockA{\textit{dept. name of organization (of Aff.)} \\
\textit{name of organization (of Aff.)}\\
City, Country \\
email address}
\and
\IEEEauthorblockN{6\textsuperscript{th} Given Name Surname}
\IEEEauthorblockA{\textit{dept. name of organization (of Aff.)} \\
\textit{name of organization (of Aff.)}\\
City, Country \\
email address}
}
\end{comment}

\IEEEoverridecommandlockouts
\IEEEpubid{\makebox[\columnwidth]{978-1-7281-2968-6/19/\$31.00~\copyright2019 IEEE \hfill} \hspace{\columnsep}\makebox[\columnwidth]{ }}

\maketitle

\IEEEpubidadjcol

\begin{abstract}
\textbf{Background:} 
Refactoring is a critical task in software maintenance and is generally performed to enforce the best design and implementation practices or to cope with design defects. Several studies attempted to detect refactoring activities through mining software repositories allowing to collect, analyze and get actionable data-driven insights about refactoring practices within software projects. 

\textbf{Aim:} 
We aim at identifying, among the various quality models presented in the literature, the ones that are more in-line
with the developer’s vision of quality optimization, when they explicitly mention that they are refactoring to improve them. 

\textbf{Method:} 
We extract a large corpus of design-related refactoring activities that are applied and documented by developers during their daily changes from 3,795 curated open source Java projects. In particular, we extract a large-scale corpus of structural metrics and anti-pattern enhancement changes, 
from which we identify 1,245 quality improvement commits with their corresponding refactoring operations, as perceived by software engineers. Thereafter, we empirically analyze the impact of these refactoring operations on a set of common state-of-the-art design quality metrics. 

\textbf{Results:}
The statistical analysis of the obtained results shows that
(i) a few state-of-the-art metrics are more popular than others; and
(ii) some metrics are being more emphasized than others.         

\textbf{Conclusions:}                                    We verify that there are a variety of structural metrics that can represent the internal quality attributes with different degrees of improvement and degradation of software quality. Most of the metrics that are mapped to the main quality attributes do capture developer intentions of quality improvement reported in the commit messages, but for some quality attributes, they don’t.
%Refactoring is a critical task in software maintenance and is generally performed to enforce best design and implementation practices, or to cope with design defects. Several studies attempted to detect refactoring activities through mining software repositories allowing to collect, analyze and get actionable data-driven insights about refactoring practices within software projects. In this paper, we extract a large corpus of design-related refactoring activities that are applied and documented by developers during their daily changes. In particular, we extract a large-scale corpus of structural metrics and anti-pattern improvement changes, from which we identify 1,245 quality improvement commits with their corresponding refactoring operations, as perceived by software engineers. Thereafter, we empirically analyze the impact of these refactoring operations on a set of common state-of-the-art design quality metrics. The statistical analysis of the obtained results show that (\textit{i}) few state-of-the-art metrics are more popular than others; and (\textit{ii}) some metrics are being more emphasized more than others.

%; and (\textit{iii}) refactoring operations, when applied in a sequence fashion, achieve better software quality improvements compared to individual refactorings.
\end{abstract}


\section{Introduction}
\label{sec:Introduction}

% Scope
Being the \textit{de facto} practice of improving software design without altering its external behavior, refactoring has been the focus on several studies, which aim to support its application by identifying refactoring opportunities, in the source code, through the optimization of structural metrics, and the removal of code smells \cite{mkaouer2014recommendation,silva2014recommending,mkaouer2014high,mkaouer2015many,couto2018quality,terra2018jmove,ubayashi2018can}. Therefore, several studies have been analyzing the impact of refactoring on existing literature quality attributes, structural metrics, and code smells \cite{alshayeb2009empirical, shatnawi2011empirical,bavota2015experimental, chavez2017does, cedrim2016does,moser2007case,wilking2007empirical,hegedHus2010effect}. The spectrum of quality attributes, structural metrics and code smells, represents the main driver for studies aiming to imitate the human decision making, and automate the refactoring process.
% Problem Statement

Despite the growing effort in recommending refactorings through structural metrics optimization and code smells removal, there is very little evidence on whether developers follow that intention when refactoring their code. A recent study by Pantiuchina et al. \cite{pantiuchina2018improving} has shown that there is a misperception between the state-of-the-art structural metrics, widely used as indicators for refactoring, and what developers actually consider to be an improvement in their source code. Thus, there is a need to distinguish, among all the structural metrics, typically used in refactoring literature, the particular ones that are of a better representation of the developers' perception of software quality improvement. 

This paper aims in identifying, among the various quality models presented in the literature, the ones that are more in-line with the developer's vision of quality, when they explicitly state that they are refactoring to improve it. 

We start with reviewing literature studies, which propose software quality attributes and their corresponding measurement in the source code, in terms of metrics. Software quality attributes are typically characterized by high-level definitions whose interpretations allow the possibility for multiple ways to calculate them in the source code. Thus, there is little consensus on what would be the optimal match between quality attributes, and code-level design metrics. For instance, as shown later in Section \ref{sec:RelatedWork}, the notion of complexity was the subject of many studies that proposed several metrics to calculate it. Therefore, we investigate which code-level metrics are more representative to the high-level quality attributes, when their optimization is explicitly stated by the developer, when applying refactorings. %Furthermore, we investigate the top performed refactoring operations, for each explicitly mentioned quality attribute. In contrary with previous studies \cite{alshayeb2009empirical, shatnawi2011empirical,bavota2015experimental, chavez2017does, cedrim2016does,moser2007case,wilking2007empirical,hegedHus2010effect}, which analyzed the impact of each individual refactoring type on structural metrics, we extract instead \textit{refactoring patterns}, i.e, top sequence of operations, as applied by developers, and found to be effective in significantly optimizing the source code, with respect to each quality attribute.

Practically, we have classified 1,245 commits, as quality improvement commits, by manually analyzing their messages and identifying an explicit statement of improving an internal quality attribute, along with detecting their refactoring activities. We mined these commits from 3,795 well-engineered, open-source projects. We identify their refactoring operations by applying state-of-the-art refactoring mining tools \cite{silva2017refdiff,tsantalis2018accurate}. We refine our dataset by untangling each commit to select only refactored code elements. Then, we cluster commits  per quality attribute (complexity, inheritance, etc.). Afterward, for each quality attribute, we calculate the values of its corresponding structural metrics, in the files, before and after their refactorings. And finally, we empirically compare the variation of these values, to distinguish the metrics that are significantly impacted by the refactorings, and so they better reflect the developer's intention of enhancing its corresponding quality attribute. To the best of our knowledge, no previous study has investigated the relationship between quality attributes and their corresponding structural metrics, from the developer's perception. Our key findings show that not all state of the art structural metrics equally represent internal quality attributes; some quality attributes are being more emphasized than others by developers. This paper extends the existing knowledge of empirically exploring the relationship between refactoring and quality as follows:
\begin{enumerate}

\item We extensively review the literature of quality attributes, used in the literature of software quality, and their corresponding possible measurements, in terms of metrics. Then we mine a large scale dataset from GitHub that consists of 1,245 commits from 3,795 software projects, proven to contain refactoring operations, and illustrating developers self-stated intentions to enhance our studied quality attributes.

\item For each quality attribute, we empirically investigate which metrics are most impacted by refactorings, and so, the closest to capture the developer's intention.

%\item For each quality attribute, we extract the top combined refactorings, applied by developers, to optimize it.

\item For reproducibility and extension, we provide a dataset of commits, their refactoring operations, and their impact on several quality metrics\footnote{https://smilevo.github.io/self-affirmed-refactoring/}.
\end{enumerate}
%especially when they explicitly mention their intention of improving quality attributes.

% Furthermore, several studies have explored the impact of refactoring on software quality \cite{alshayeb2009empirical, shatnawi2011empirical,bavota2015experimental, chavez2017does, cedrim2016does,moser2007case,wilking2007empirical,hegedHus2010effect}, by analyzing the effect of individual refactoring types, on structural metrics. Yet, there is little knowledge about the impact of refactorings when applied in sequences. 

% Remainder
The remainder of this paper is organized as follows: Section \ref{sec:RelatedWork} reviews the existing studies related to measuring software quality and analyzing the relationship between quality attributes and refactoring. Section \ref{sec:EmpiricalSetup} outlines our empirical setup in terms of data collection, analysis and research questions. Section \ref{sec:ResultsDiscussion} discusses our findings, while Section \ref{sec:Threats} captures any threats to the validity of our work, before concluding with Section \ref{sec:conclusion}.
\section{Related Work}
\label{sec:RelatedWork}

%\subsection{Impact of Refactoring on Quality}
\begin{table*}
  \centering
	 \caption{A summary of the literature on the impact of refactoring activities on software quality attributes.}
	 \label{Table:Quality Metrics in Related Work}
%\begin{sideways}
\begin{adjustbox}{width=1.0\textwidth,center}
%\begin{adjustbox}{width=\textheight,totalheight=\textwidth,keepaspectratio}
\begin{tabular}{lllllll}\hline
\toprule
\bfseries Study & \bfseries Year & \bfseries Approach & \bfseries Software Metric & \bfseries Internal QA & \bfseries External QA  \\
\midrule
Sahraoui et al. \cite{sahraoui2000can} & 2000 & Analyzing code histories & CLD / NOC / NMO / NMI    & Inheritance / Coupling & Fault-proneness / Maintainability   \\ 
& & & NMA / SIX / CBO / DAC & & \\
& & & IH-ICP / OCAIC / DMMEC / OMMEC &  \\ \hline
Stroulia \& Kapoor \cite{stroulia2001metrics} & 2001 & Performing a case study & LOC / LCOM / CC & Size / Coupling & Design extensibility \\ \hline
Kataoka et al. \cite{kataoka2002quantitative} & 2002 & Analyzing code histories & Coupling measures &  Coupling & Maintainability   \\ \hline
Demeyer \cite{demeyer2002maintainability} & 2002 & Analyzing code histories & N/A & Polymorphism  & Performance  \\ \hline
Tahvildari et al. \cite{tahvildari2003quality} & 2003 & Analyzing code histories & LOC / CC / CMT / Halstead's efforts & Complexity  & Performance / Maintainability   \\ \hline
Leitch \& Stroulia \cite{leitch2003assessing}& 2003 & Analyzing code histories & SLOC / No. of Procedure & Size & Maintainability  \\ \hline
Bois \& Mens \cite{du2003describing} & 2003 & Analyzing code histories & NOM / CC / NOC / CBO & Inheritance / Cohesion / Coupling / Size / Complexity & N/A  \\ 
& & & RFC / LCOM & & \\ \hline
Tahvildari \& Kontogiannis \cite{tahvildari2003metric} & 2004 & Analyzing code histories & LCOM / WMC / RFC / NOM   & Inheritance / Cohesion / Coupling / Complexity & Maintainability & \\ 
& & & CDE / DAC / TCC & & \\ \hline
Bois et al. \cite{du2004refactoring} & 2004 & Analyzing code histories & N/A & Cohesion / Coupling & Maintainability   \\ \hline
Bois et al. \cite{du2005does} & 2005 & Analyzing code histories & N/A & N/A &   Understandability   \\  \hline
Geppert et al. \cite{geppert2005refactoring} & 2005 & Performing a case study &  N/A & N/A & Changeability  \\ \hline
Ratzinger et al. \cite{ratzinger2005improving} & 2005 & Mining commit log & N/A &  Coupling & Evolvability \\ 
& & Analyzing code histories & \\ \hline
Moser et al. \cite{moser2006does} & 2006 & Analyzing code histories & CK / MCC / LOC   & Inheritance / Cohesion / Coupling / Complexity & Reusability \\ \hline
Wilking et al. \cite{wilking2007empirical} & 2007 & Analyzing code histories & CC / LOC  & Complexity & Maintainability / Modifiability   \\ \hline
Stroggylos \& Spinells \cite{stroggylos2007refactoring} & 2007 & Mining commit log & CK / Ca / NPM & Inheritance / Cohesion / Coupling / Complexity & N/A  \\ \hline
Moser et al. \cite{moser2007case} & 2008 & Analyzing code histories & CK / LOC / Effort (hour) & Cohesion / Coupling / Complexity & Productivity  \\ \hline
Alshayeb \cite{alshayeb2009empirical} & 2009 & Analyzing code histories &  CK / LOC / FANOUT  & Inheritance / Cohesion / Coupling / Size & Adaptability / Maintainability / Testability / Reusability  \\ 
& & & & & Understandability  \\ \hline
Hegedus et al. \cite{hegedHus2010effect} & 2010 & Analyzing code histories & CK  & Coupling / Complexity / Size & Maintainability / Testability / Error Proneness / Changeability  \\
& & & & & Stability / Analizability &\\ \hline
Shatnawi \& Li \cite{shatnawi2011empirical} & 2011 & Analyzing code histories & CK / QMOOD &  Inheritance / Cohesion / Coupling / Polymorphism / Size & Reusability / Flexibility / Extendibility / Effectiveness     \\ 
& & & & Encapsulation / Composition / Abstraction / Messaging  & &  \\ \hline
Bavota et al. \cite{bavota2013empirical} & 2013 & Analyzing code histories & ICP / IC-CD / CCBC & Coupling & N/A \\
& & Surveying developers & & & \\ \hline
Szoke et al. \cite{szoke2014bulk} & 2014 & Mining commit log & CC / U / NOA / NII / NAni & Size / Complexity & N/A \\
& & Surveying developers & LOC / NUMPAR / NMni / NA & &  \\ \hline
Bavota et al. \cite{bavota2015experimental} & 2015 & Mining commit log &  CK / LOC / NOA / NOO  &  Inheritance / Cohesion / Coupling / Size / Complexity & N/A \\
& & Analyzing code histories & C3 / CCBC & & \\ \hline
Cedrim at al. \cite{cedrim2016does} & 2016 & Mining commit log & LOC / CBO / NOM / CC & Cohesion / Coupling / Complexity & N/A  \\
& & Analyzing code histories & FANOUT / FANIN & &  \\ \hline
Chavez et al. \cite{chavez2017does} & 2017 & Mining commit log & CBO / WMC / DIT / NOC & Inheritance / Cohesion / Coupling / Size / Complexity & N/A  \\ 
& & Analyzing code histories & LOC / LCOM2 / LCOM3 / WOC & &  \\
& & & TCC / FANIN / FANOUT / CINT & &  \\
& & & CDISP / CC / Evg / NPATH   & & \\
& & & MaxNest / IFANIN / OR / CLOC & &  \\
& & & STMTC / CDL / NIV / NIM / NOPA & &  \\ \hline 
Pantiuchina et al. \cite{pantiuchina2018improving} & 2018 & Mining commit log & LCOM / CBO / WMC / RFC  & Cohesion / Coupling / Complexity & Readability  \\
&  & Analyzing code histories &  C3 / B\&W / SRead & \\

\bottomrule
\end{tabular}
\end{adjustbox}
%\end{sideways}
\end{table*}
It is widely acknowledged in the literature of software refactoring that it has the ultimate goal to improve software quality and fix design and implementation bad practices \cite{Fowler:1999:RID:311424}. %\ali{some references, e.g. fowler, etc.}.
In recent year, there is much research efforts have focused on studying and exploring the impact of refactoring on software quality \cite{moser2007case,wilking2007empirical,alshayeb2009empirical, shatnawi2011empirical,bavota2015experimental, chavez2017does,mkaouer2017robust, cedrim2016does,hegedHus2010effect}. 
%\ali{some references}.
The vast majority of studies have focused on measuring the internal and external quality attributes to determine the overall quality of a software system being refactored. In this section, we review and discuss the relevant literature on the impact of refactoring on software quality.

In an academic setting, Stroulia and Kapoor \cite{stroulia2001metrics} investigate the effect of size and coupling measures on software quality after the application of refactoring. The results in Stroulia and Kapoor's work show that size and coupling metrics decreased after refactorings.  Kataoka et al. \cite{kataoka2002quantitative} used only coupling measures to study the impact of \textit{Extract Method} and \textit{Extract Class} refactoring operations on the maintainability of a single C++ software system, and found that refactoring has positive impact on system maintainability. Demeyer \cite{demeyer2002maintainability} performed a comparative study to investigate the impact of refactoring on performance. The results of Demeyer's study show that program performance is enhanced after the application of refactoring. Moreover, Sahraoui et al. \cite{sahraoui2000can} used coupling and inheritance measures to automatically detect potential anti-patterns and predict situations where refactoring could be applied to improve software maintainability. The authors found that quality metrics can help to bridge the gap between design improvement and its automation, but in some situations the process cannot be fully automated as it requires the programmer's validation through manual inspection. 

Tahvildari et al. \cite{tahvildari2003quality} proposed a software transformation framework that links software quality requirements like performance and maintainability with program transformation to improve the target qualities. The results show that utilizing design patterns increase system's maintainability and performance. In another study, Tahvildari and Kontogiannis \cite{tahvildari2003metric} used the same framework to evaluate four object-oriented measures (\textit{i.e.,} cohesion, coupling, complexity, and inheritance) in addition to software maintainability. Leitch and Stroulia \cite{leitch2003assessing} used dependency graph-based techniques to study the impact of two refactorings, namely, \textit{Extract Method} and \textit{Move Method}, on software maintenance using two small systems. The authors found that refactoring enhanced the quality by (1) reducing the design size, (2) increasing number of procedures, (3) reducing the data dependencies, and (4) reducing regression testing. Bios and Mens \cite{du2003describing} proposed a framework to analyze the impact of three refactorings on five internal quality attributes (\textit{i.e.,} cohesion, coupling, complexity, inheritance, and size), and their findings show positive and negative impacts on the selected measures. Bios et al. \cite{du2004refactoring} provided a set of guidelines for optimizing cohesion and coupling measures. This study shows that the impact of refactoring on these measures ranged from negative to positive. In a follow-up work,  Bios et al. \cite{du2005does} conducted a study to differentiate between the application of Refactor to Understand and the traditional Read to Understand pattern. Their findings show that refactoring plays a role in improving the understandability of the software. 

Geppert et al. \cite{geppert2005refactoring} investigated the impact of refactoring on changeability focusing on three factors for changeability, namely, customer-reported defect rates, change effort, and scope of changes. Their findings show a significant decrease in the first two factors. Ratzinger et al. \cite{ratzinger2005improving} analyzed the historical data of a large industrial system and focused on reducing change couplings. Based on the identified change couplings, they also analyzed code smell changes for the purpose of identifying where to apply refactoring efficiently. They concluded that refactoring is able to enhance software evolvability (\textit{i.e.,} reduce the change coupling). In an agile development environment, Moser et al. \cite{moser2006does} used internal measures (\textit{i.e.,} CK, MCC, LOC) to explore the effect of refactoring on the reusability of the code using a commercial system, and found that refactoring was able to improve the reusability of hard-to-reuse classes. Wilking et al. \cite{wilking2007empirical} empirically studied the effect of refactoring on non-functional aspects, \textit{i.e.,} the maintainability and  modifiability of system systems. They tested the maintainability by explicitly adding defects to the code, and then they measured the time taken to remove them. Modifiability, on the other hand, was examined by adding new functionalities and then measuring the LOC metric and the time taken to implement these features. The authors did not find a clear effect of refactoring on these two external attributes. 

Stroggylos and Spinellis \cite{stroggylos2007refactoring} opted for
searching words stemming from the verb ``refactor" such
as \say{refactoring} or \say{refactored} to identify refactoring-related commits to study the impact of refactoring on quality using eight object-oriented metrics. Their results indicated possible negative effects of refactoring on quality, \textit{e.g.,} increased LCOM metric. Moser et al. \cite{moser2007case} studied the impact of refactoring on the productivity in an agile team. The achieved results show that refactoring improved software developers' productivity besides several aspects of quality, \textit{e.g.}, maintainability. Alshayeb \cite{alshayeb2009empirical} conducted a study aiming at assessing the impact of eight refactorings on five external quality attributes (\textit{i.e.,} adaptability, maintainability, understandability, reusability, and testability). The author found that refactoring could improve the quality in some classes, but could also decrease software quality to some extent in other classes. Hegedus et al. \cite{hegedHus2010effect} examined the effect of singular refactoring techniques on testability, error proneness, and other maintainability attributes. They concluded that refactoring could have undesired side effects that can degrade the quality of the source code. 

In an empirical setting, Shatnawi and Li \cite{shatnawi2011empirical} used the hierarchical quality model to assess the impact of refactoring on four software quality factors, namely, reusability, flexibility, extendibility, and effectiveness. The authors found that the majority of refactoring operations exhibit positive impact on quality; however, some operations deteriorated quality. Bavota et al. empirically investigated the developers' perception of coupling, as captured by structural, dynamic, semantic, and logical coupling measures. They found that semantic coupling measure aligns with developers' perceptions better that the other coupling measures. In a more recent study, Bavota et al. \cite{bavota2015experimental} used RefFinder\footnote{https://github.com/SEAL-UCLA/Ref-Finder}, 
%\cite{5609577}
a version-based refactoring detection tool, to mine the evolution history of three open-source systems. They mainly investigated the relationship between refactoring and quality. The study findings indicate that 42\% of the performed refactorings are affected by code smells, and refactorings were able to eliminate code smells in only 7\% of the cases. 

Cedrim et al. \cite{cedrim2016does} conducted a longitudinal study of 25 projects to investigate the improvement of software structural quality. They analyzed the relationship of refactorings and code smells by classifying refactorings according to the addition or removal of poor code structures. The study results indicate that only 2.24\% of refactorings removed code smells, and 2.66\% introduced new ones. Recently, Chavez et al. \cite{chavez2017does} studied the effect of refactoring on five internal quality attributes, namely, cohesion, coupling, complexity, inheritance, and size, using 25 quality metrics. The study shows that root-canal refactoring-related operations are either improved or at least not worsened the internal quality attributes. Additionally, when floss refactoring-related operations are applied, 55\% of these operations improved these attributes, while only 10\% of quality declined. 

In particular, two studies \cite{szoke2014bulk,pantiuchina2018improving} are most related to our work have analyzed the comment commits in which developers stated the purpose of improving the quality. Szoke et al. \cite{szoke2014bulk} studied 198 refactoring commits of five large-scale industrial systems to investigate the effects of these commits on quality of several revisions for a period of time. To know the purpose of the applied refactorings, they trained developers and asked them to state the reason when committing the changes to the repositories, which could be related to (1) fix coding issues, (2) fix anti-patterns, and (3) improve certain metrics. The study results show that performing a single refactoring could negatively impact the quality, but applying refactorings in blocks (\textit{e.g.,} fixing more coding issues or improving more quality metrics) can significantly improve software quality. More recently, Pantiuchina et al. \cite{pantiuchina2018improving} empirically investigated the correlation between seven code metrics and the quality improvement explicitly reported by developers in 1,282 commit messages. The study shows that quality metrics sometimes do not capture the quality improvement reported by developers. A common indicator to assess the quality improvements between these studies resides in the use the quality metrics. Both of these studies found that minor refactoring changes rarely impact the quality of the software.

All of the above-mentioned studies have focused on assessing the impact of refactorings on the quality by either considering the internal or the external quality attributes using a variety of approaches. Among them, few studies \cite{ratzinger2005improving, stroggylos2007refactoring, szoke2014bulk, bavota2015experimental, cedrim2016does, chavez2017does, pantiuchina2018improving} mined software repositories to explore the impact on quality. Otherwise, the vast majority of these studies %except Bavota et al. \cite{bavota2015experimental} and Szuke et al. \cite{szoke2014bulk},
used a limited set of projects and mined general commits without applying any form of verification regarding whether refactorings have actually been applied. 

Our work is different from these studies as our main purpose is to explore if there is an alignment between quality metrics and quality improvements that are documented by developers in the commit messages. As we summarize these state-of-the-art studies in Table~\ref{Table:Quality Metrics in Related Work}. We identify 8 popular quality attributes, namely \textit{Cohesion}, \textit{Coupling}, \textit{Complexity}, \textit{Inheritance}, \textit{Polymorphism}, \textit{Encapsulation}, \textit{Abstraction} and \textit{Design size}. As different studies advocate for various metrics to calculate these quality attributes, we extract and calculate 27 structural metrics. In particular, on a more qualitative sense, we conduct an empirical study using 1,245 commits that are proven to contain real-world instances of refactoring activities, in the purpose of improving software design. To the best of our knowledge, no previous study has empirically investigated, using a curated set of commits, the representativeness of structural design metrics for internal quality attributes. In the next section, we detail the steps we took to design our empirical setup.
%while considering real-world instances of refactoring commits applied by developers and their commit message as identified by Refactoring Miner \cite{tsantalis2018accurate}.% to contain at least one refactoring operation. 

%\ali{Great job for the related work section! but quite detailed for a conference paper :-)}
\section{Empirical Study Setup}
\label{sec:EmpiricalSetup}

%This section presents our empirical study. 
Our main goal is to investigate whether the developer perception of quality improvement (as expected by developers) aligns with the real %quantitative assessment 
quality improvement (as assessed by quality metrics).  %\ali{just added parenthesis for better explanation}
In particular, we address the following research question:

\begin{itemize}
\item \textit{Is the developer perception of quality improvement aligned with the quantitative assessment of code quality?}

%What are the structural metrics that better capture the developer's impact when explicitly optimizing internal quality attributes?} 

\begin{table}
  \centering
	 \caption{Internal quality attributes and their corresponding structural metrics used in this study.}
	 \label{Table:Quality Metrics Used in This Study.}
%\begin{sideways}
%\begin{adjustbox}{width=1.1\textwidth,center}
%\begin{adjustbox}{width=\textheight,totalheight=\textwidth,keepaspectratio}
\begin{tabular}{lll}\hline
\toprule
\bfseries Quality Attribute & \bfseries Study & \bfseries Software Metrics \\
\midrule
%\multicolumn{2}{l}{\textbf{\textit{Internal Quality Attribute }}}\\
%\midrule
Cohesion & \cite{pantiuchina2018improving,chavez2017does} & Lack of Cohesion of Methods (LCOM) \cite{chidamber1994metrics}  \\ 
Coupling &  \cite{chavez2017does,pantiuchina2018improving} & Coupling Between Objects (CBO) \cite{chidamber1994metrics}  \\
         & \cite{pantiuchina2018improving} & Response For Class (RFC) \cite{chidamber1994metrics}  \\
         & \cite{chavez2017does} & Fan-in (FANIN) \cite{henry1981software} \\     
         & \cite{chavez2017does} & Fan-out (FANOUT) \cite{henry1981software} \\
Complexity & \cite{chavez2017does} & Cyclomatic Complexity (CC) 
           \cite{mccabe1976complexity}\\
           & \cite{chavez2017does,pantiuchina2018improving,singh2012evaluation} & Weighted Method Count (WMC) \cite{chidamber1994metrics}  \\
           & \cite{neelamegam2009survey,singh2012evaluation} & Response For Class (RFC) \cite{chidamber1994metrics}  \\
           & \cite{singh2012evaluation} & Lack of Cohesion of Methods (LCOM) \cite{chidamber1994metrics} \\
           & \cite{chavez2017does} & Essential Complexity (Evg) \cite{mccabe1976complexity}\\
           & \cite{chavez2017does} & Paths (NPATH) \cite{nejmeh1988npath} \\
           & \cite{chavez2017does} & Nesting (MaxNest)
           \cite{lorenz1994object} \\
Inheritance & \cite{chavez2017does,singh2012evaluation} & Depth of Inheritance Tree (DIT) \cite{chidamber1994metrics}  \\
            & \cite{chavez2017does,singh2012evaluation} & Number of Children (NOC) \cite{chidamber1994metrics} \\
            & \cite{chavez2017does} & Base Classes (IFANIN) \cite{Destefanis:2014:SMA:2813544.2813555} \\
Polymorphism & \cite{singh2012evaluation} & Weighted Method Count (WMC) \cite{chidamber1994metrics}  \\
             & \cite{neelamegam2009survey,singh2012evaluation} & Response For a Class (RFC) \cite{chidamber1994metrics}  \\
Encapsulation & \cite{singh2012evaluation} & Weighted Method Count (WMC) \cite{chidamber1994metrics} \\
& \cite{singh2012evaluation} & Lack of Cohesion of Methods (LCOM) \cite{chidamber1994metrics} \\
Abstraction & \cite{singh2012evaluation} & Weighted Method Count (WMC) \cite{chidamber1994metrics} \\
& \cite{singh2012evaluation} & Lack of Cohesion of Methods (LCOM) \cite{chidamber1994metrics} \\ 
Design Size & \cite{chavez2017does} & Lines of Code (LOC) \cite{lorenz1994object} \\
            & \cite{chavez2017does} & Lines with Comments (CLOC) \cite{lorenz1994object}  \\
            & \cite{chavez2017does} & Statements (STMTC) \cite{lorenz1994object} \\
            & \cite{chavez2017does} & Classes (CDL) \cite{lorenz1994object}\\
            & \cite{chavez2017does} & Instance Variables (NIV) \cite{lorenz1994object} \\
            & \cite{chavez2017does} & Instance Methods (NIM) \cite{lorenz1994object} \\
%Composition & \\
%\midrule 
%\multicolumn{2}{l}{\textbf{\textit{External Quality Attribute }}}\\
%\midrule
%Adabtability & \cite{dandashi2002method,alshayeb2009empirical} & Depth of Inheritance Tree (DIT) \cite{chidamber1994metrics}\\
%Maintainability &  & Number of Children (NOC) \cite{chidamber1994metrics} \\
%Understandability &  & Coupling Between Objects (CBO) \cite{chidamber1994metrics} \\
%Reusability &  & Response For a Class (RFC) \cite{chidamber1994metrics} \\
%Completeness &  & Weighted Method Count (WMC) \cite{chidamber1994metrics} \\
%  &  & Lines of Code (LOC) \cite{lorenz1994object} \\
%Testability & \cite{bruntink2006empirical,alshayeb2009empirical} &  
%Response For a Class (RFC) \cite{chidamber1994metrics} \\
%&  & Fan-out (FANOUT) \cite{henry1981software} \\
%&  & Weighted Method Count (WMC) \cite{chidamber1994metrics} \\
%&  & Lines of Code (LOC) \cite{lorenz1994object} \\
\bottomrule
\end{tabular}
%\end{adjustbox}
%\end{sideways}
\end{table}

%\item \textbf{RQ2.} \textit{What are the refactoring sequences that positively improve the internal quality attributes?} \ali{I suggest to remove the word "positively"}
\end{itemize}

%\ali{I just mentioned the RQs a bit early here to improve the flow of the paper. Feel free to remove this part if you dot not agree}\Eman{Right. I think it is good either to have them here or in the intro section}

\begin{figure*}[ht]
%\begin{sideways}
\centering 
\includegraphics[width=\textwidth]{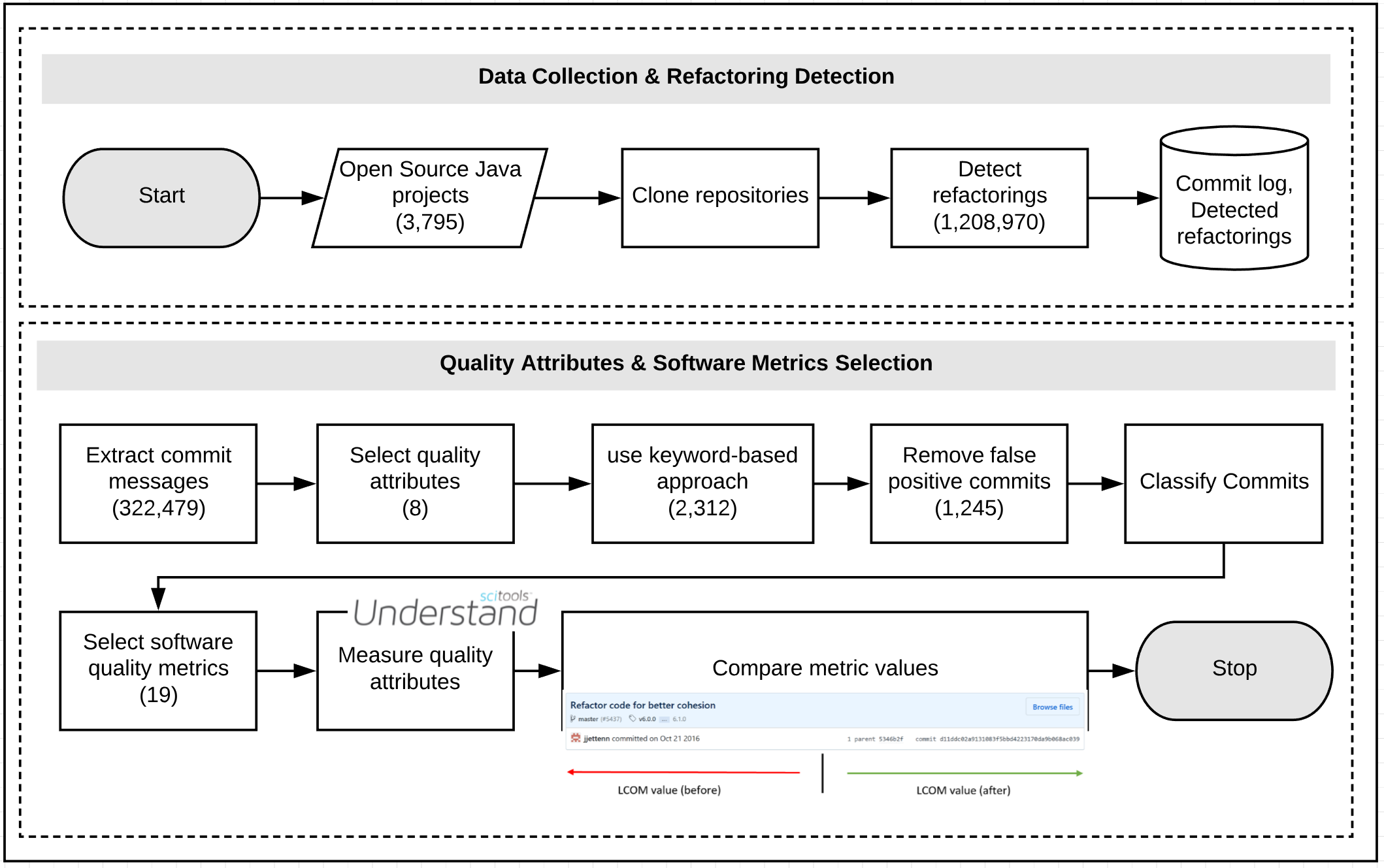}
%\includegraphics[width=\columnwidth]{Images/Methodology.PNG}
%\end{sideways}
\caption{Empirical study design overview.}
\label{fig:approach_overview}
\end{figure*}

To answer our research question, we conduct a three-phased empirical study. An overview of the experiment methodology is depicted in Figure~\ref{fig:approach_overview}. The initial phase consists of selecting and mining a large number of open-source Java projects and detecting refactoring instances that occur throughout their development history, \textit{i.e.,} commit-level code changes, of each considered project. The second phase consists of analyzing the commit messages as a mean of identifying refactoring commits in which developers document their perception of internal quality attributes. Thereafter, the third phase involves the selection of software quality metrics to compare its values before and after the selected refactoring commits. 

\subsection{Selection of Quality Attributes and Structural Metrics}
%After mapping each refactoring operation found in our golden set to internal quality attributes, we map quality metrics to each internal quality attribute, similar to \cite{chavez2017does}. 
To setup a comprehensive set of quality attributes, to be assessed in our study, we first conduct a literature review on existing and commonly acknowledged software quality attributes \cite{chidamber1994metrics,lorenz1994object,mccabe1976complexity, henry1981software, nejmeh1988npath, Destefanis:2014:SMA:2813544.2813555}. Then, we checked if the metrics assess several object-oriented design aspects %\ali{the term "feature" seems too broad; do you mean "design aspects" or characteristics, etc.?}\Eman{yes} 
in order to map each internal quality attribute to the appropriate structural metric(s). %\ali{it could not be external too?}\Eman{in this study, we consider only the internal QAs. I have just cleaned the discussion about the external ones that we initially planned to cover}. 
For example, the Response For Class (RFC) metric is typically used to measure Coupling and Complexity quality attributes. More generally, we extract, from literature review, all the associations between metrics (\textit{e.g.,} CK suite \cite{chidamber1994metrics}, McCabe \cite{mccabe1976complexity} and Lorenz and Kidd's book \cite{lorenz1994object}) with internal quality attributes.% considered in this study. %As for the external quality attributes, we consider the software metrics defined by Dandashi \cite{dandashi2002method} and Bruntink and Deursen \cite{bruntink2006empirical} when correlating the internal software metrics with the external quality attributes. 

The extraction process results in 27 distinct structural metrics as shown in Table~\ref{Table:Quality Metrics Used in This Study.}. The list of metrics is (1) well-known and defined in the literature, and (2) can assess on different code-level elements, \textit{i.e.}, method, class, package, and (3) can be calculated by existing static analysis tools. %\ali{which tool? you never talked about this tool before. I suggest to replace it by "can be calculated by existing quality assurance tools (e.g., ...), or simply remove (3)}. 
For this study, all metrics values are automatically computed using the \textsc{Understand}\footnote{\url{https://scitools.com/}}, a popular static analysis framework. %\ali{Ah, now you mention the tool :-)}

\subsection{Refactoring Detection}

To collect the necessary commits, we refer to an existing large dataset of links to GitHub repositories \cite{allamanis2013mining}. We perform an initial filtering, using Reaper \cite{munaiah2017curating}, to only navigate through well-engineered projects. So, we ended up reducing the number of selected projects from 57,447 to 3,795. To extract the entire refactoring history in each project, we use two popular refactoring mining tools, namely Refactoring Miner \cite{silva2016we}  and ReffDiff \cite{silva2017refdiff}. We selected both tools because they are known to be in the top of refactoring detection tools, in terms of accuracy \cite{tsantalis2018accurate,tan2019survey} (precision of 98\% and 100\%, and recall of 87\% and 88\%, respectively), and because they are both built-in to analyze code changes in git repositories and detect applied refactorings, which is the case for our intended data, along with being suitable for our study that requires a high degree of automation in data mining. %Our choice to use Refactoring Miner is justified by the fact that it achieved the highest accuracy in detecting refactorings compared to the state-of-the-art available tools, with a precision of 98\% and recall of 87\% \cite{silva2016we,tsantalis2018accurate}. Refactoring Miner seems %The Eclipse plug-in refactoring detection tools require user interaction to select projects as inputs and trigger the refactoring detection, which is impractical since multiple releases of the same project have to be imported to Eclipse to identify the refactoring history. 
As for the selection of commits with refactorings, we perform a voting process between both tools, \textit{i.e.,} in order for a given commit to be selected, it has to be detected by both tools as a container to at least one refactoring operation. We perform this voting process to raise the likelihood of refactoring existence in the commit. Since the accuracy of the tools is out of the scope of this work, and since we do not perform any refactoring-related analysis, we do not care if the detection results overlap or not.

In this phase, We collect a total of 1,208,970 refactoring operations from 322,479 commits, applied during a period of 23 years (1997-2019). An overview of the studied benchmark is provided in Table~\ref{Table:DATA_Overview}.

\begin{table}[h]
\begin{center}
\caption{Studied dataset statistics.}
\label{Table:DATA_Overview}
\begin{tabular}{lr}\hline
\toprule
\bfseries Item & \bfseries Count \\
\midrule
Studied projects & 3,795 \\
%All commits & \\
%Commits filtered by mining tools & \\
Commits with refactorings & 322,479 \\
Refactoring operations & 1,208,970 \\
Commits with refactorings \& Keywords & 2,312 \\
Remove false positive commits & 1,067 \\
Final dataset & 1,245\\

%\midrule 
%\multicolumn{2}{c}{\textbf{\textit{Analyzed Projects - Refactored Code Elements}}}\\
%\bfseries Code Element & \bfseries \# of Refactorings  \\
%\midrule
%Class & 329,378  \\
%Method & 718,335 \\
%Attribute & 97,516 \\
%Package & 18,334 \\
%Interface & 8,096 \\
\bottomrule
\end{tabular}
\end{center}
\end{table}

\subsection{Data Extraction}
After extracting all refactoring commits, we want to only keep commits where refactoring is documented i.e., self-affirmed refactorings \cite{alomar2019empirical}. We continue to filter them, using the content of their messages at this stage. We start with using a keyword-based search to find commits whose messages contain one of the keywords (\textit{i.e., Cohesion, Coupling, Complexity, Inheritance, Polymorphism, Encapsulation, Abstraction, size})%related to our internal quality attributes. 
%as well as a subset of external qualities which are correlated with internal software metrics as defined by Dandashi and Bruntink \& Deursen \cite{dandashi2002method, bruntink2006empirical}. Dandashi \cite{dandashi2002method} assessed five indirect quality attributes (\textit{i.e.,} Adabtability, Maintainability, Understandability, Reusability, and Completeness) from the direct quality attributes. Bruntink and Deursen \cite{bruntink2006empirical} also used a set of internal quality metrics as indicators of Testability attribute. 

\begin{table*}[h]
\begin{center}
\caption{Examples of selected commit messages.}
\label{Table:Commit Message Examples}
\begin{adjustbox}{width=1.0\textwidth,center}
\begin{tabular}{ll}\hline
\toprule
\bfseries Quality Attribute & \bfseries Commit Message \\
\midrule
Cohesion & \textit{Refactor code for better cohesion}%\tablefootnote{\url{https://github.com/molgenis/molgenis/commit/d11ddc02a9131083f5bbd4223170da9b068ac039}}
\\
Coupling & \textit{Reduce coupling between packages}%\tablefootnote{\url{https://github.com/veithen/visualwas/commit/f4bae91a0a163317aa157fff2718d3e837191fb4}}
\\
Complexity & \textit{reducing complexity by refactoring}%\tablefootnote{\url{https://github.com/deegree/deegree3/commit/d402daf08289e3145e9999f65e0226fd41ff9e09}}
\\
Inheritance & \textit{refactored document requests code to better reflect inheritance ...}%\tablefootnote{\url{https://github.com/FenixEdu/fenixedu-academic/commit/e0d30084d88b503fa1ee06bbd43c9b039dc7d433}} 
\\
Polymorphism & \textit{Enhance field manager to account for polymorphism when getting a field from a ceiling class}%\tablefootnote{\url{https://github.com/BroadleafCommerce/BroadleafCommerce/commit/8ad5ba0c2609c82552399d9ee922a792a1214025}}
\\
Encapsulation & \textit{Refactored transactional observer code for better encapsulation and runtime performance} %\tablefootnote{\url{https://github.com/weld/core/commit/8addb072a5192c7caf6b404b0c3015c48e83d0b1}}
\\
Abstraction & \textit{code refactored in order to improve the abstraction}%\tablefootnote{\url{https://github.com/antoniomaria/gazpachoquest/commit/ec4accb8eb35a10b5ca45957214ec2d980a52efa}} 
\\
Design Size & \textit{Major refactoring to reduce code size and have at least halfway reasonable structure ... }%\tablefootnote{\url{https://github.com/wocommunity/wonder/commit/a2408ac6e3f195842bd68b5a0f96098bc0c6241d}}  
\\
\bottomrule
\end{tabular}
\end{adjustbox}
\end{center}
\end{table*}

This keyword-based filtering resulted in only selecting 2,312 commit messages. We notice that the ratio of these commits is very small in comparison with the total number of refactoring commits, \textit{i.e.,} 322,479. However, these observations are aligned with previous studies \cite{murphy2012we,szoke2014bulk} as developers typically do not provide details when they document their refactorings. To ensure that these commits reported developers' intention to improve quality attributes, we manually inspect and read through these refactoring commits to remove false positives. An example of a discarded commit is: \say{\textit{Refactored EphemeralFileSystemAbstraction}}. We discarded this commit because the quality attribute is actually part of the identifier name of the class. In case of disagreement between the authors on the inclusion of a certain commit, it was excluded. This step resulted in only considering 1,245 commits. During this process, we manually classified them with respect to their quality attributes, as one commit could belong to more than one quality attribute. Our goal is to have a \textit{gold set} of commits in which the developers explicitly reported the quality attributes improvement. This \textit{gold set} will serve to check later if there is an alignment between the real quality metrics affected in the source code, and the quality improvement as documented by developers. Examples of commit messages belonging to the \textit{gold set}, are showcased in Table \ref{Table:Commit Message Examples}.

Since commits typically contain multiples changed files, which may not all be involved in the refactoring, we filter them out, as we checkout, for each commit, its changed Java files, and keep only those involved in the refactoring operation(s), associated with that commit. %impacted by the refactorings (\textit{i.e.,} only refactored files) before and after the changes implemented by refactoring commits.
The resulting commits, correspond to our data points, each data point is represented by a set of \textit{pre-refactoring} and \textit{post-refactoring} Java files. These data points will be used in the experiments, to measure the effect of changes in terms of structural metrics, with respect to the quality attribute, announced in the commit message.
%\subsection{Mapping Refactoring to Quality Attributes}
%Similar to \cite{chavez2017does}, we map each refactoring operation to the corresponding quality attribute as reported in the related works \cite{Fowler:1999:RID:311424,kerievsky2005refactoring} to better understand the impact of refactoring operations on quality. Table depicts the association of each refactoring type with the internal quality attributes that are expected to be improved.

\section{Empirical Study Results \& Discussion}
\label{sec:ResultsDiscussion}

\begin{figure*}
% row-1
\centering
\begin{subfigure}{4.5cm}
\centering\includegraphics[width=4cm]{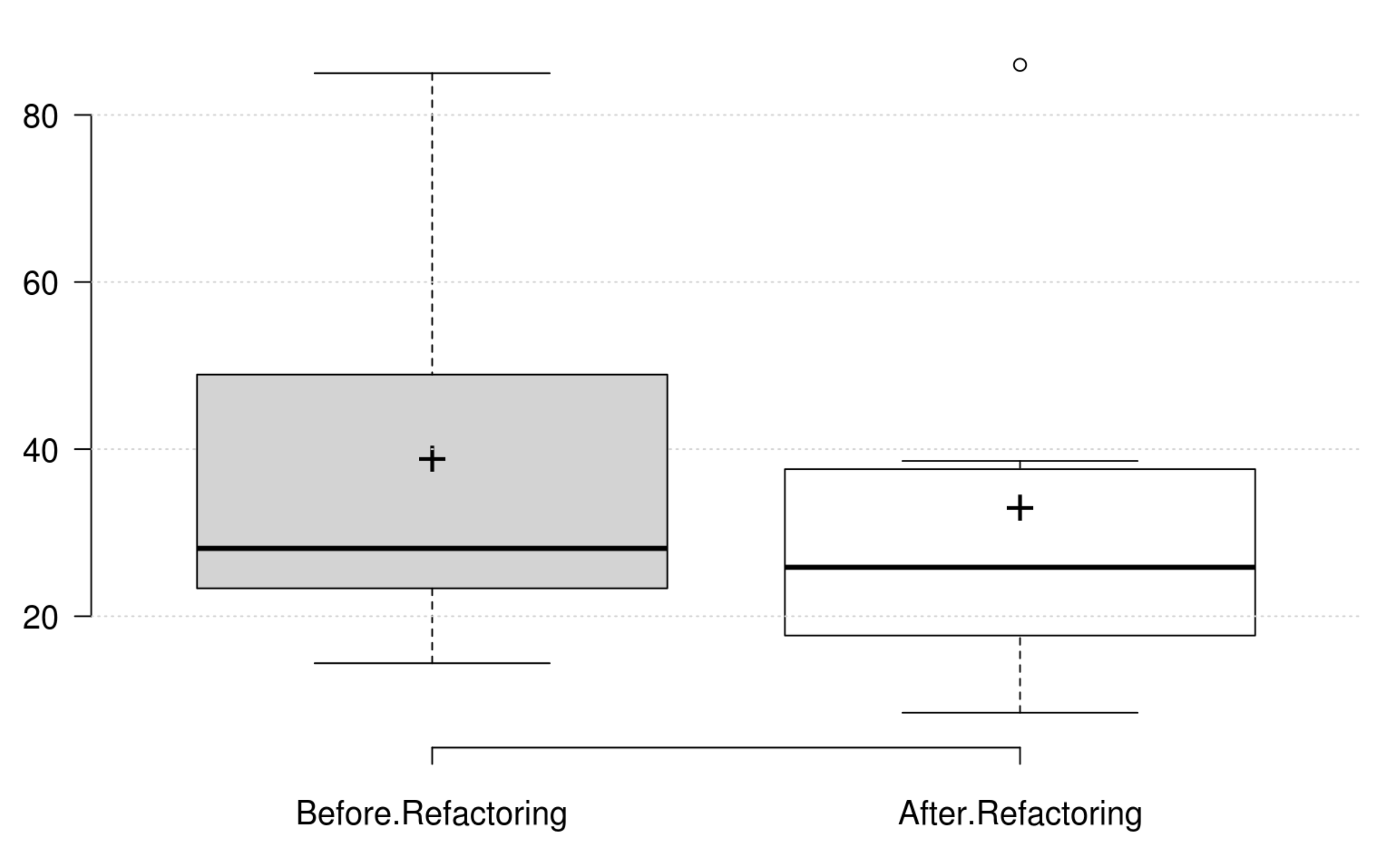}
\caption{Cohesion - LCOM}
\label{BP:chesion-lcom}
\end{subfigure}%
\begin{subfigure}{4.5cm}
\centering\includegraphics[width=4cm]{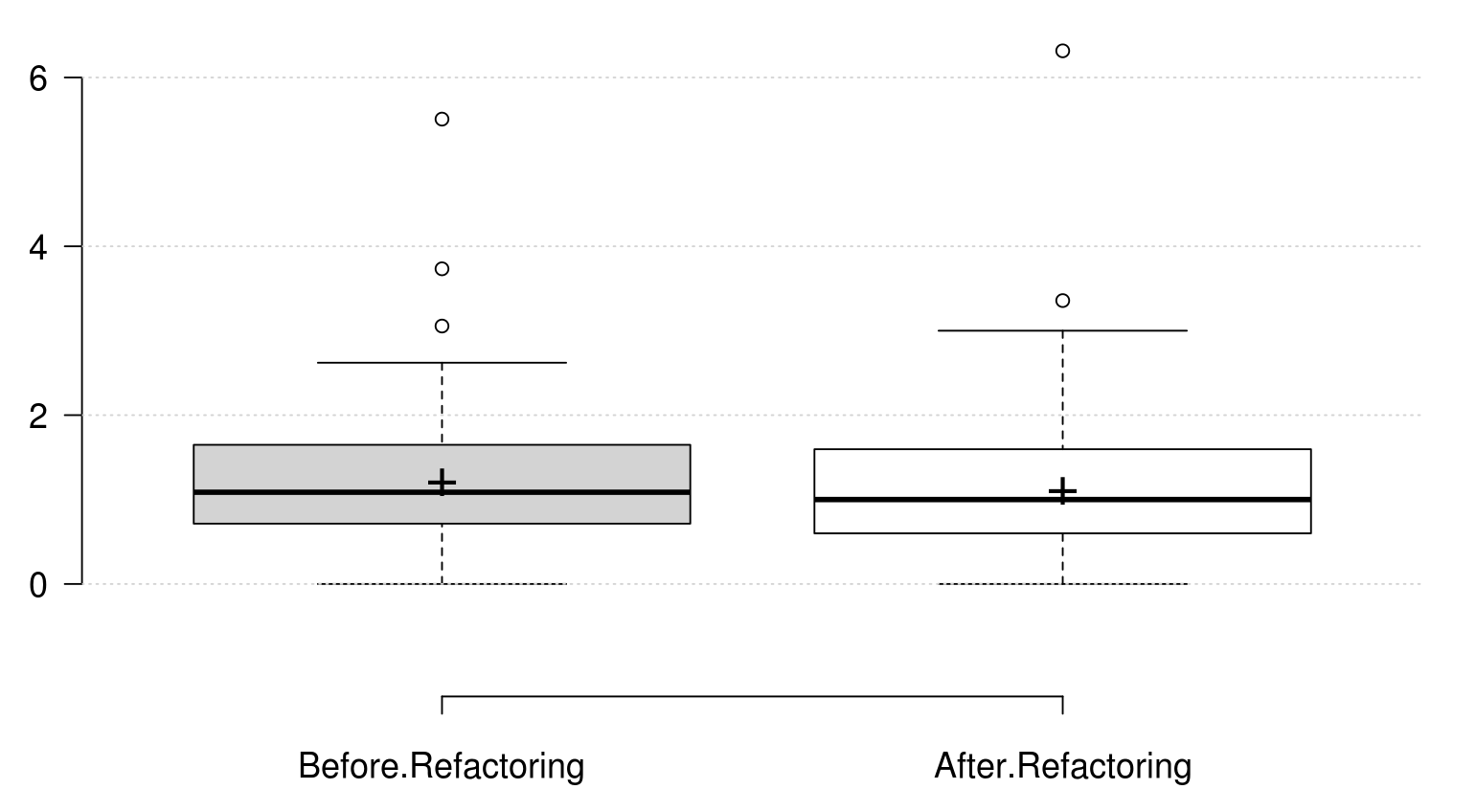}
\caption{Coupling - CBO}
\label{BP:coupling-cbo}
\end{subfigure}%
\begin{subfigure}{4.5cm}
\centering\includegraphics[width=4cm]{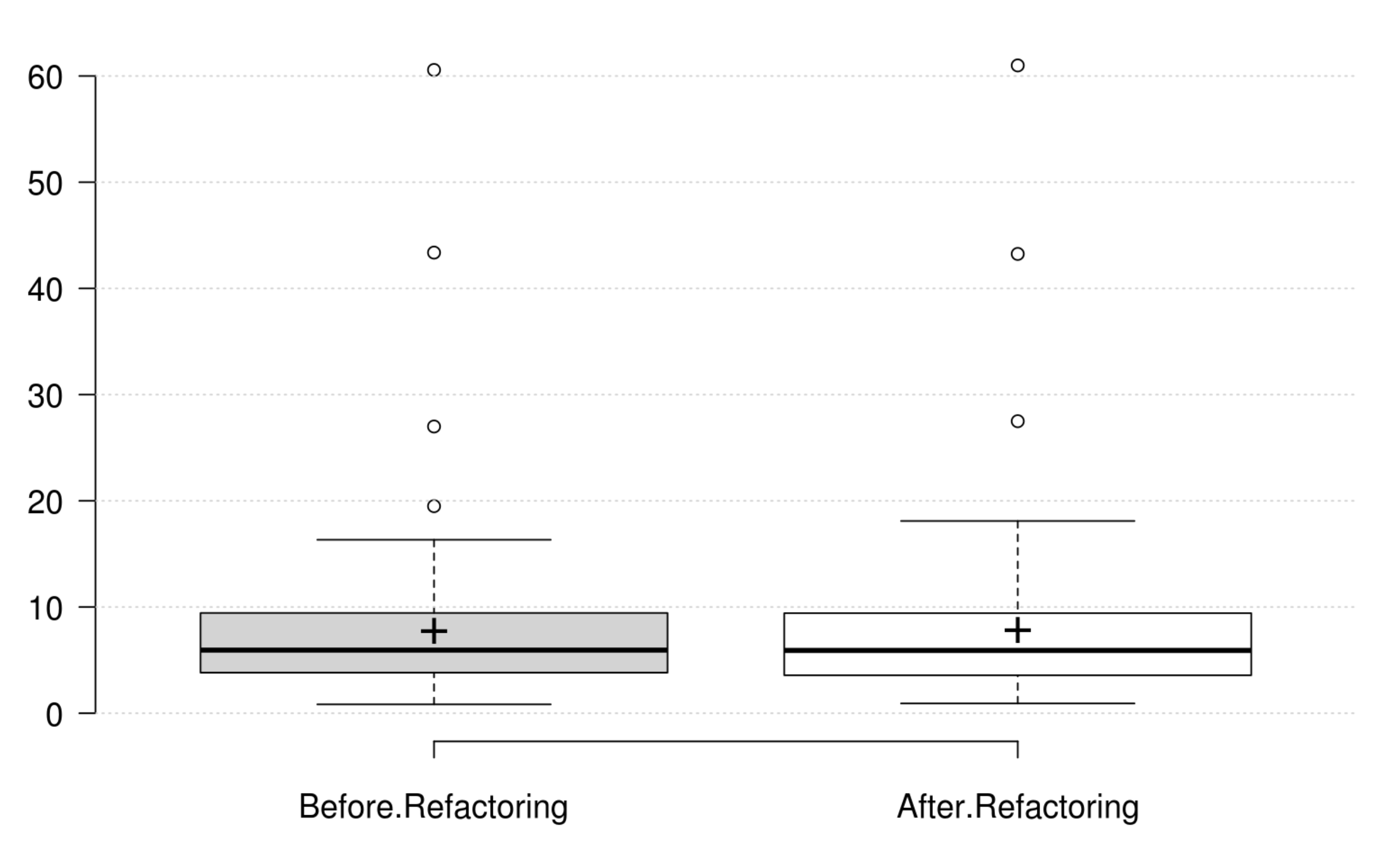}
\caption{Coupling - FANIN}
\label{BP:coupling-rfc}
\end{subfigure}%\vspace{9pt}
\begin{subfigure}{4.5cm}
\centering\includegraphics[width=4cm]{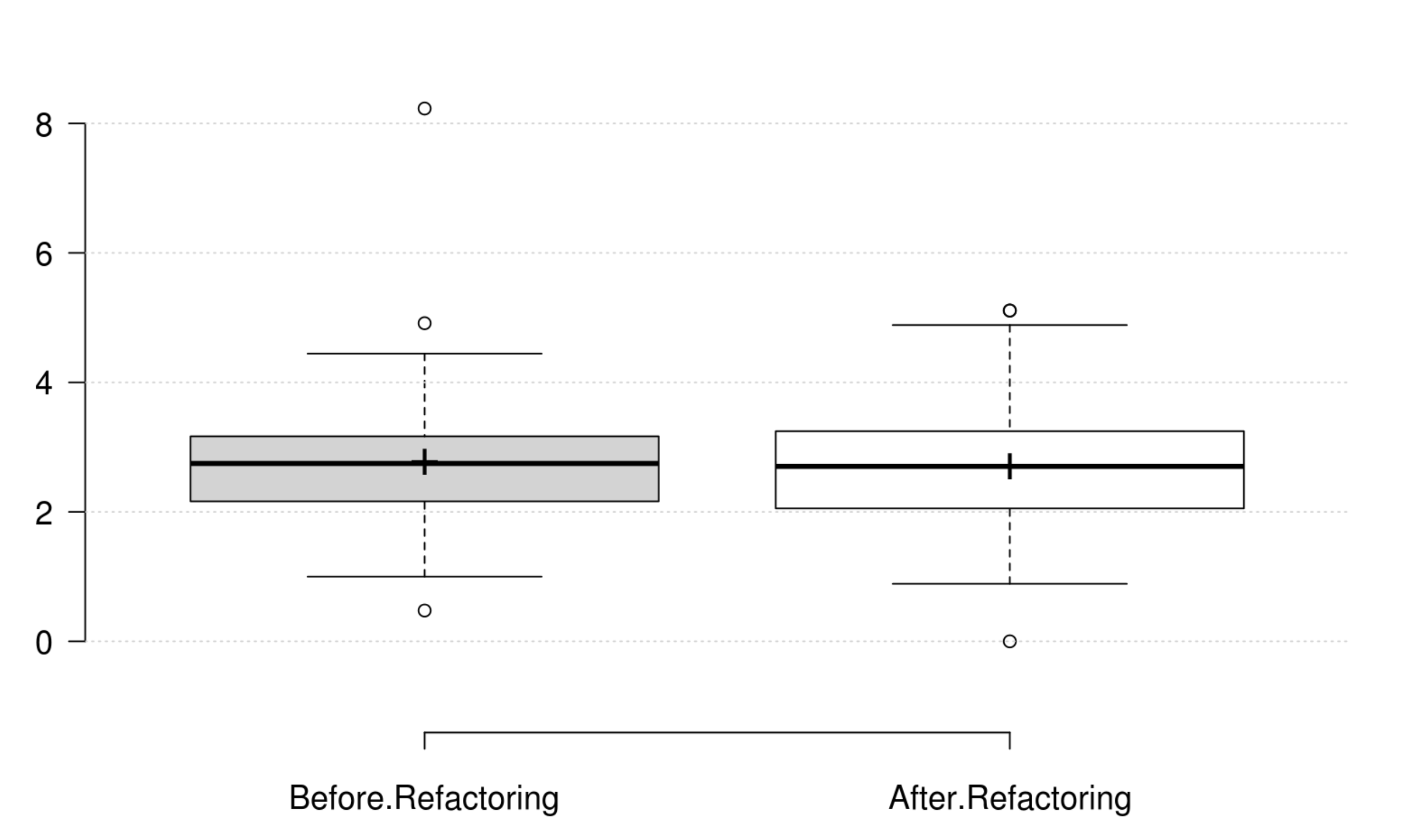}
\caption{Coupling - FANOUT}
\label{BP:coupling-fanin}
\end{subfigure}%

% row-2 
\begin{subfigure}{4.5cm}
\centering\includegraphics[width=4cm]{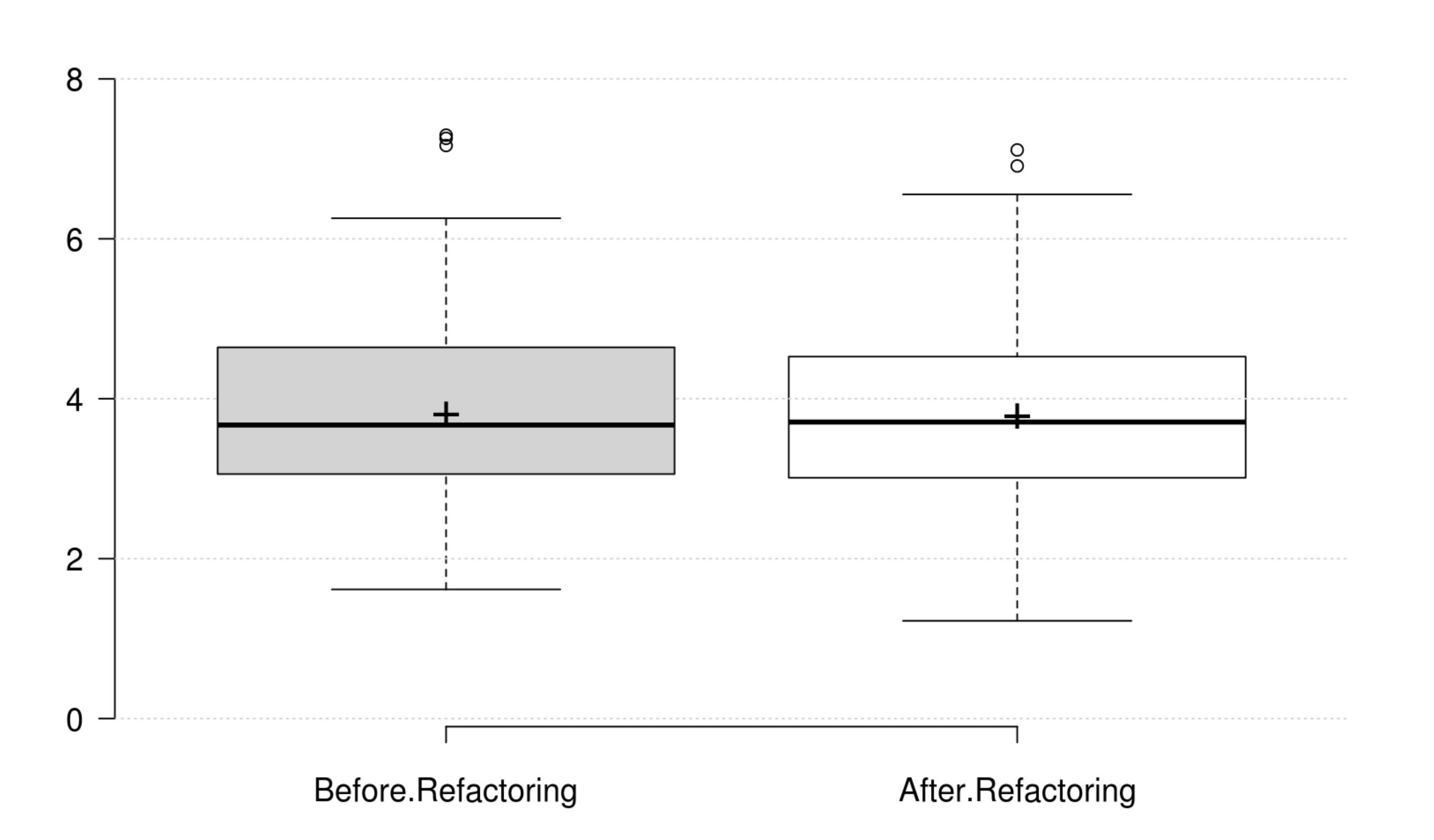}
\caption{Coupling - RFC}
\label{BP:coupling-fanout}
\end{subfigure}%
\begin{subfigure}{4.5cm}
\centering\includegraphics[width=4cm]{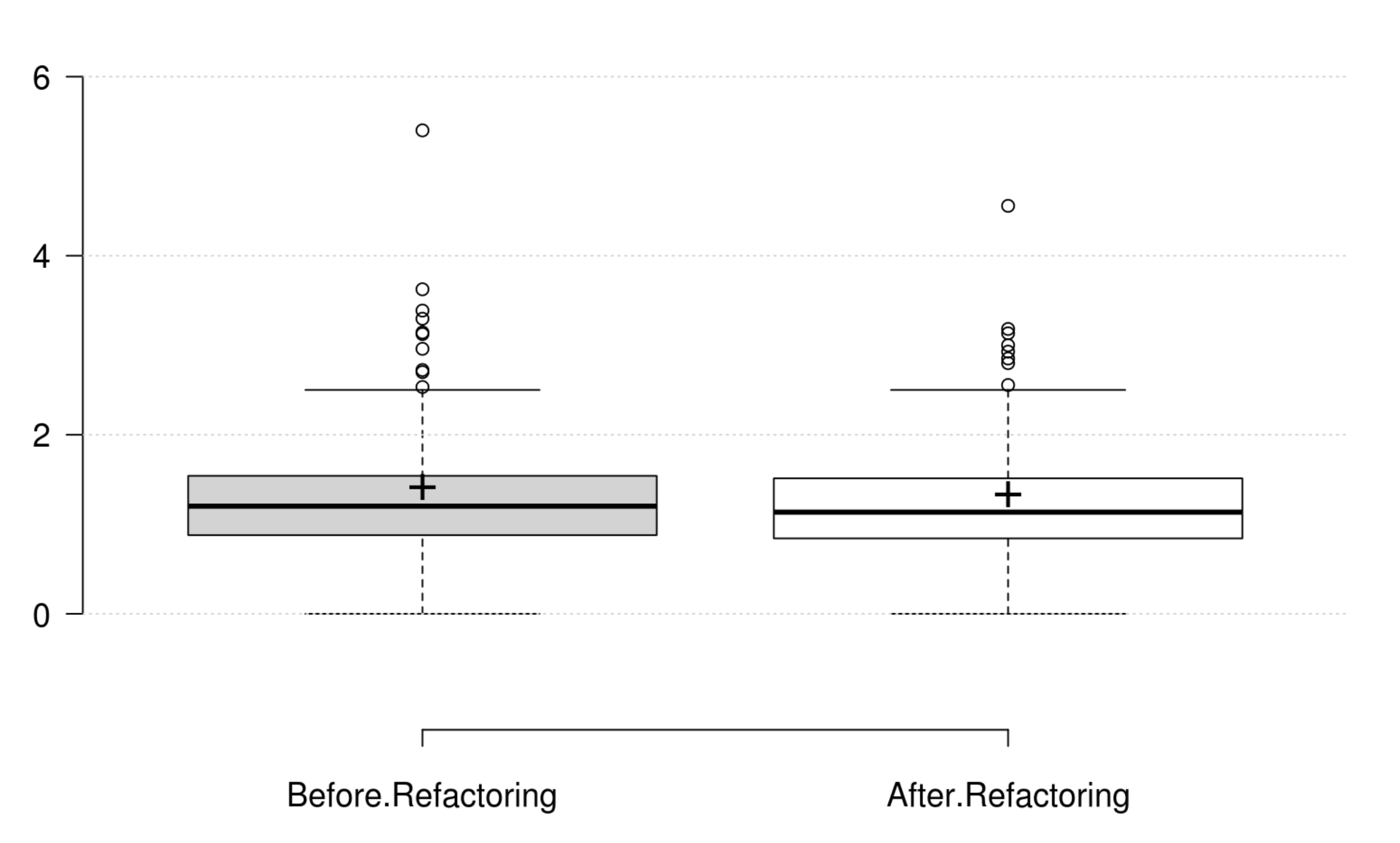}
\caption{Complexity - CC}
\label{BP:complexity-cc}
\end{subfigure}%\vspace{9pt}
\begin{subfigure}{4.5cm}
\centering\includegraphics[width=4cm]{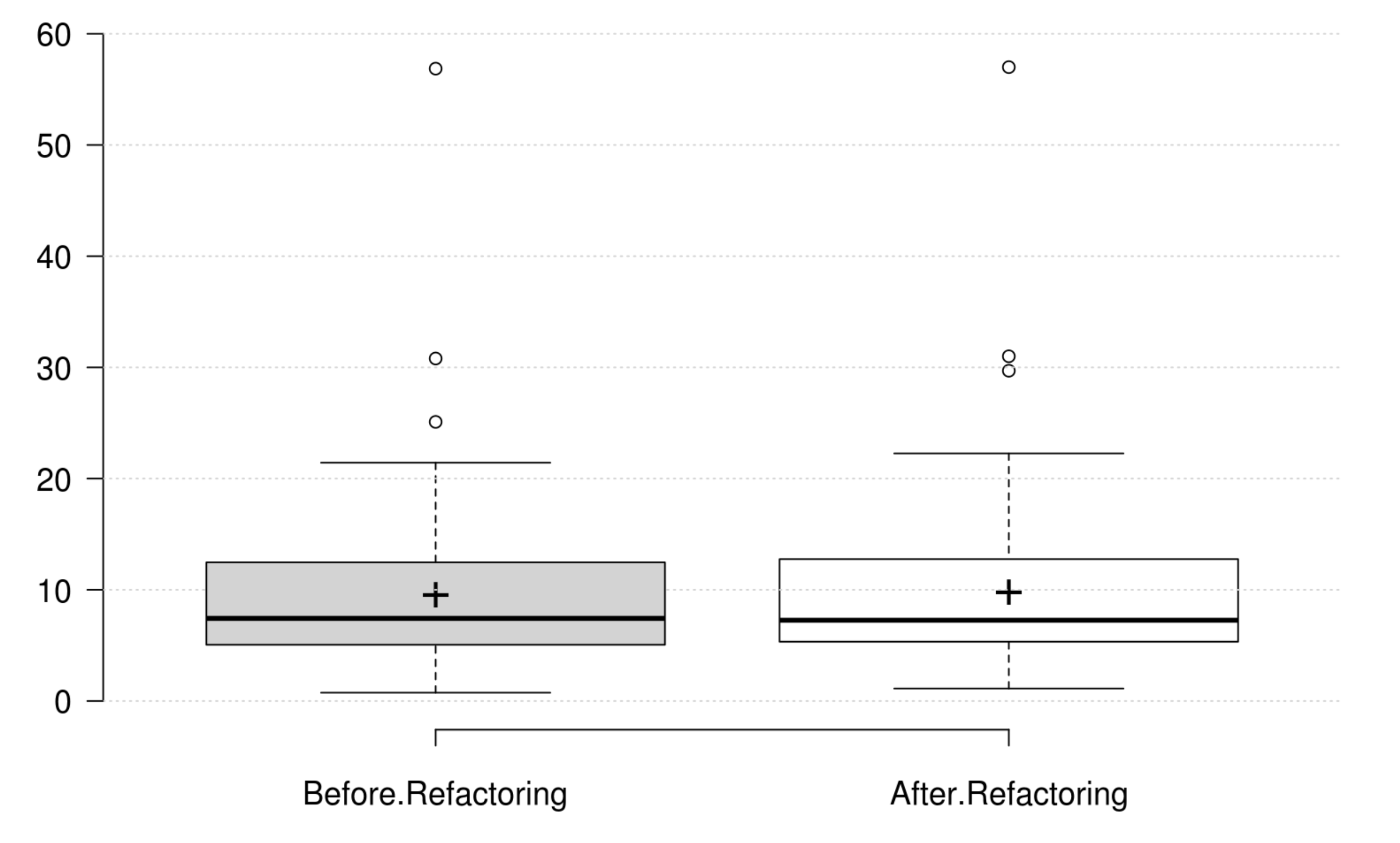}
\caption{Complexity - WMC}
\label{BP:complexity-wmc}
\end{subfigure}%
\begin{subfigure}{4.5cm}
\centering\includegraphics[width=4cm]{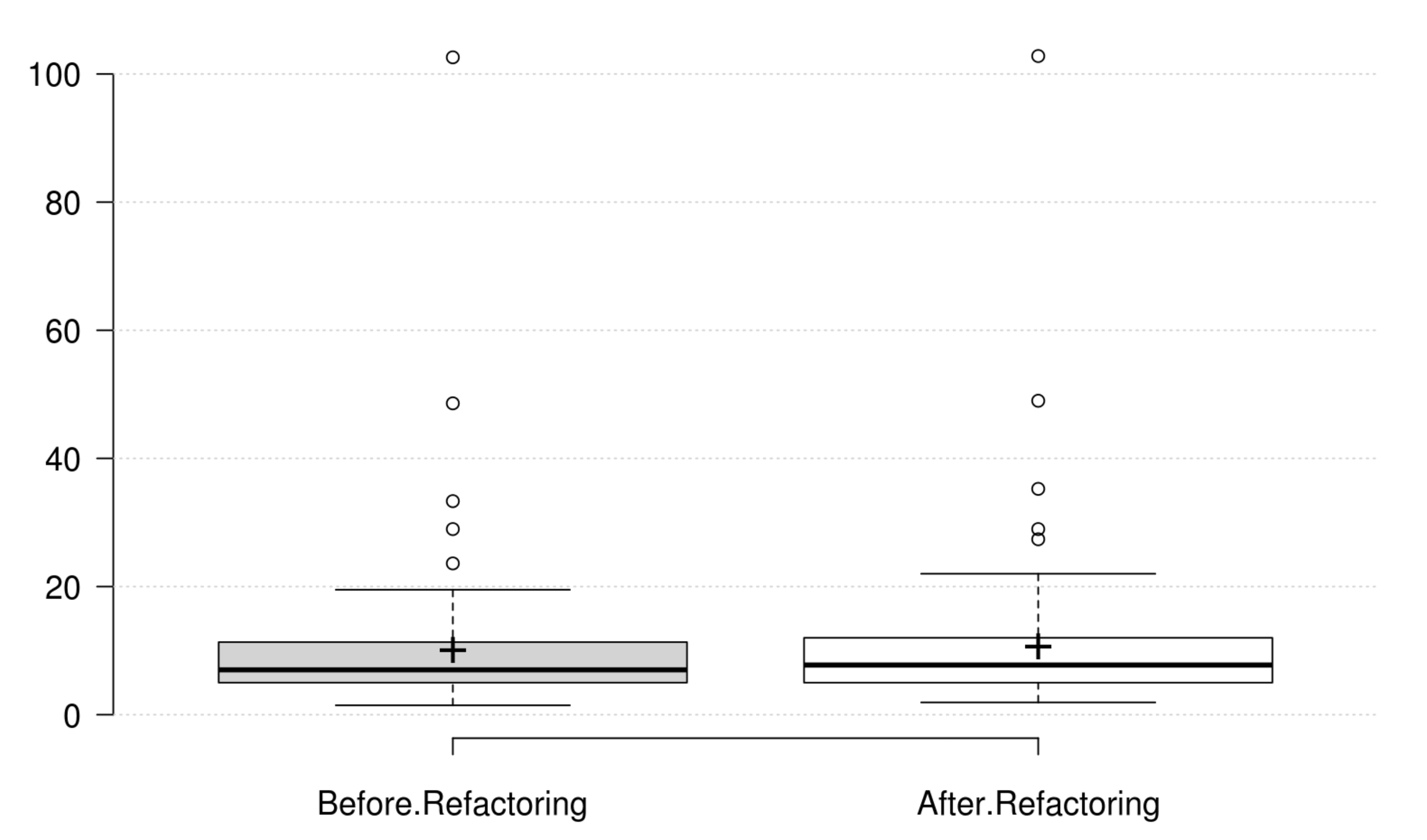}
\caption{Complexity - RFC}
\label{BP:complexity-rfc}
\end{subfigure}%

% row-3
\begin{subfigure}{4.5cm}
\centering\includegraphics[width=4cm]{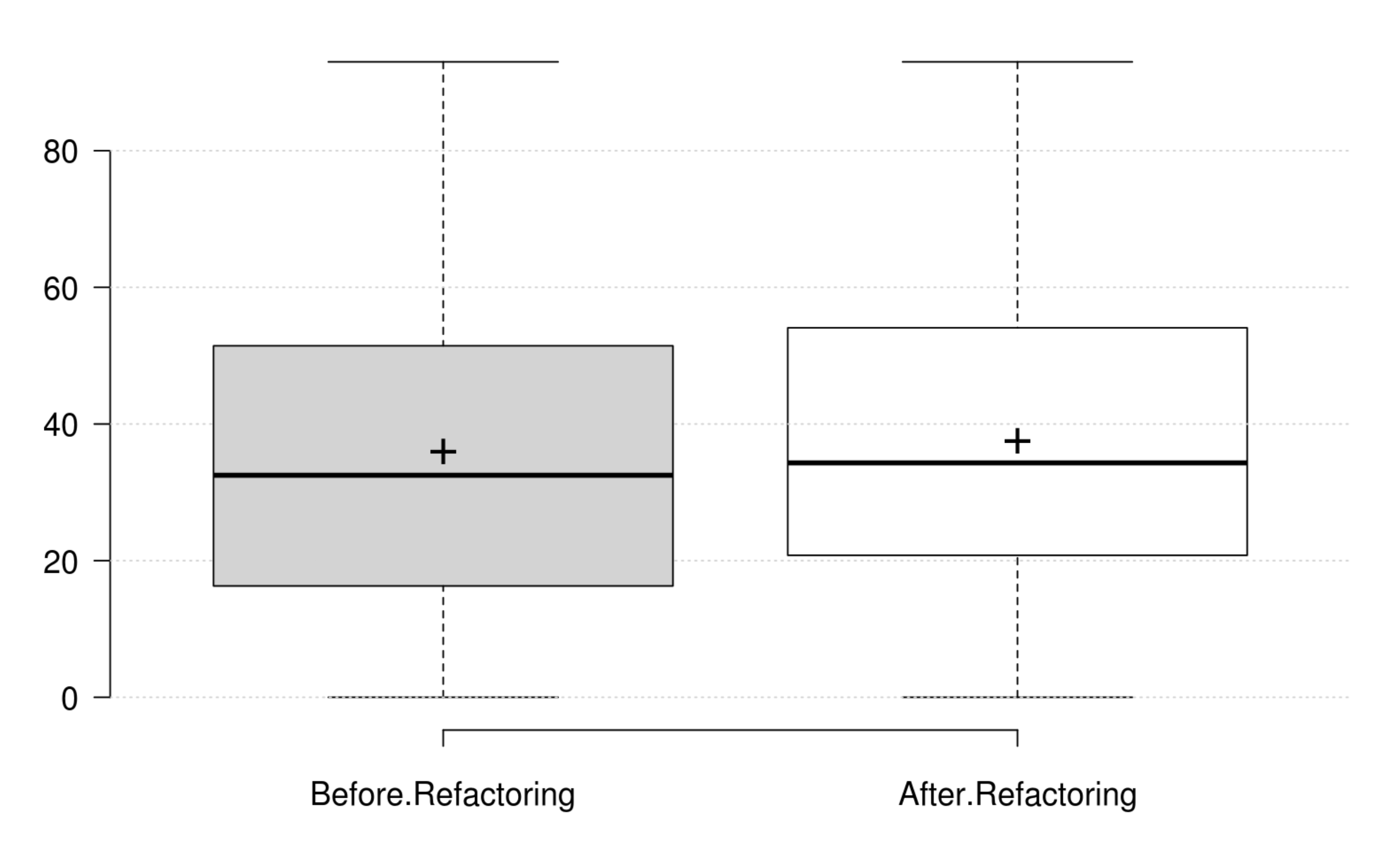}
\caption{Complexity - LCOM}
\label{BP:complexity-lcom}
\end{subfigure}%
\begin{subfigure}{4.5cm}
\centering\includegraphics[width=4cm]{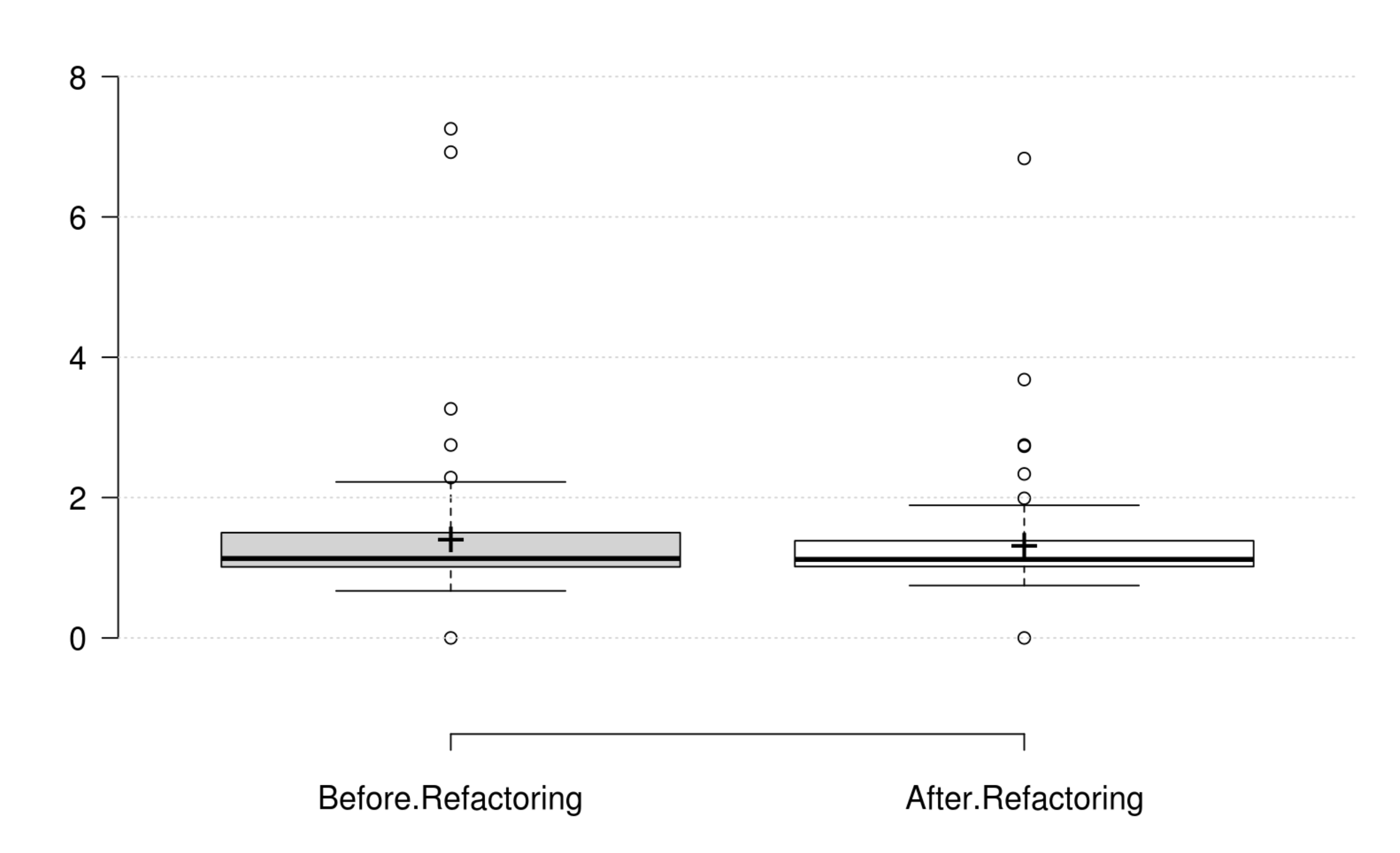}
\caption{Complexity - Evg}
\label{BP:complexity-evg}
\end{subfigure}
\begin{subfigure}{4.5cm}
\centering\includegraphics[width=4cm]{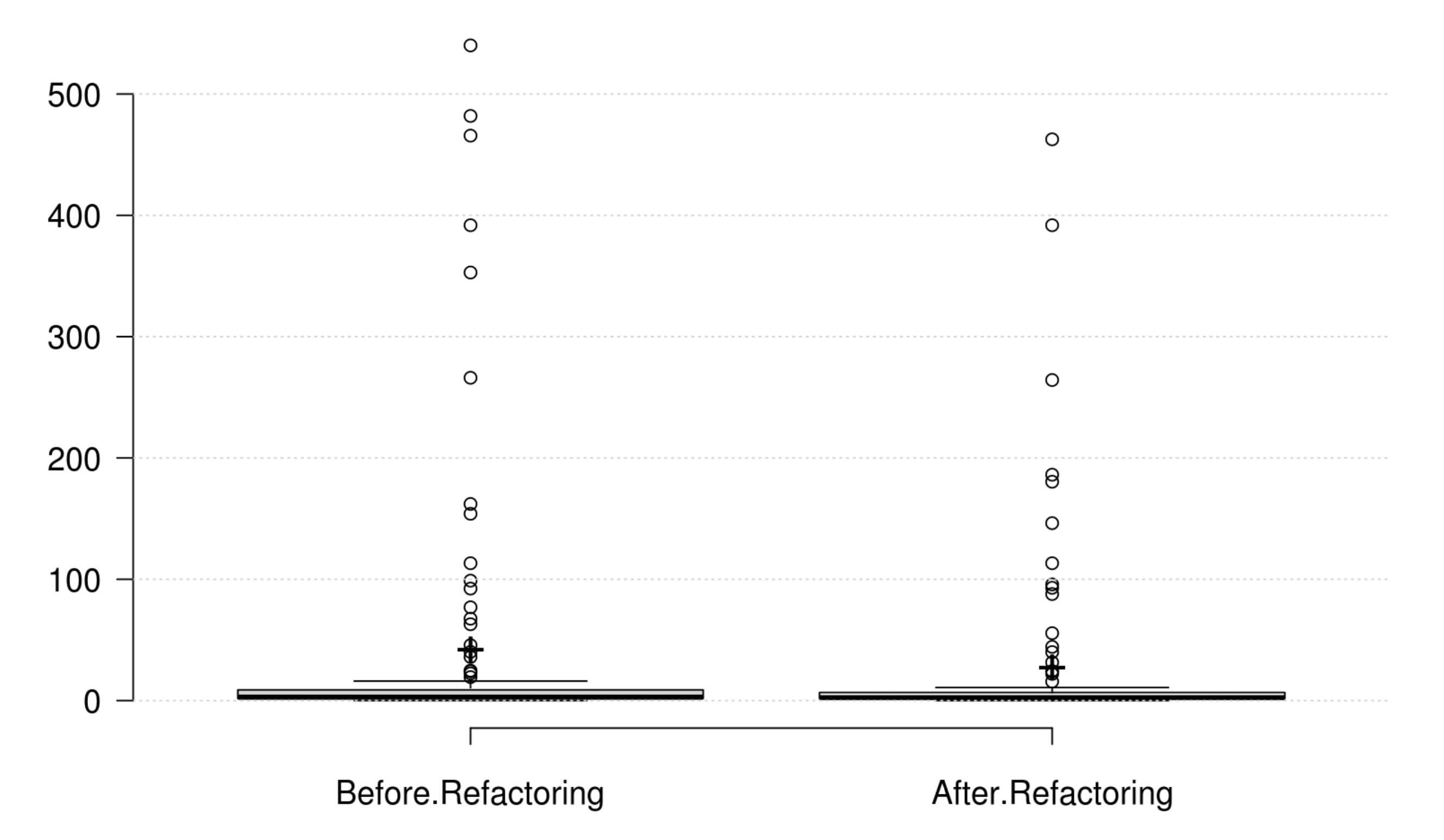}
\caption{Complexity - NPATH}
\label{BP:complexity-npath}
\end{subfigure}%
\begin{subfigure}{4.5cm}
\centering\includegraphics[width=4cm]{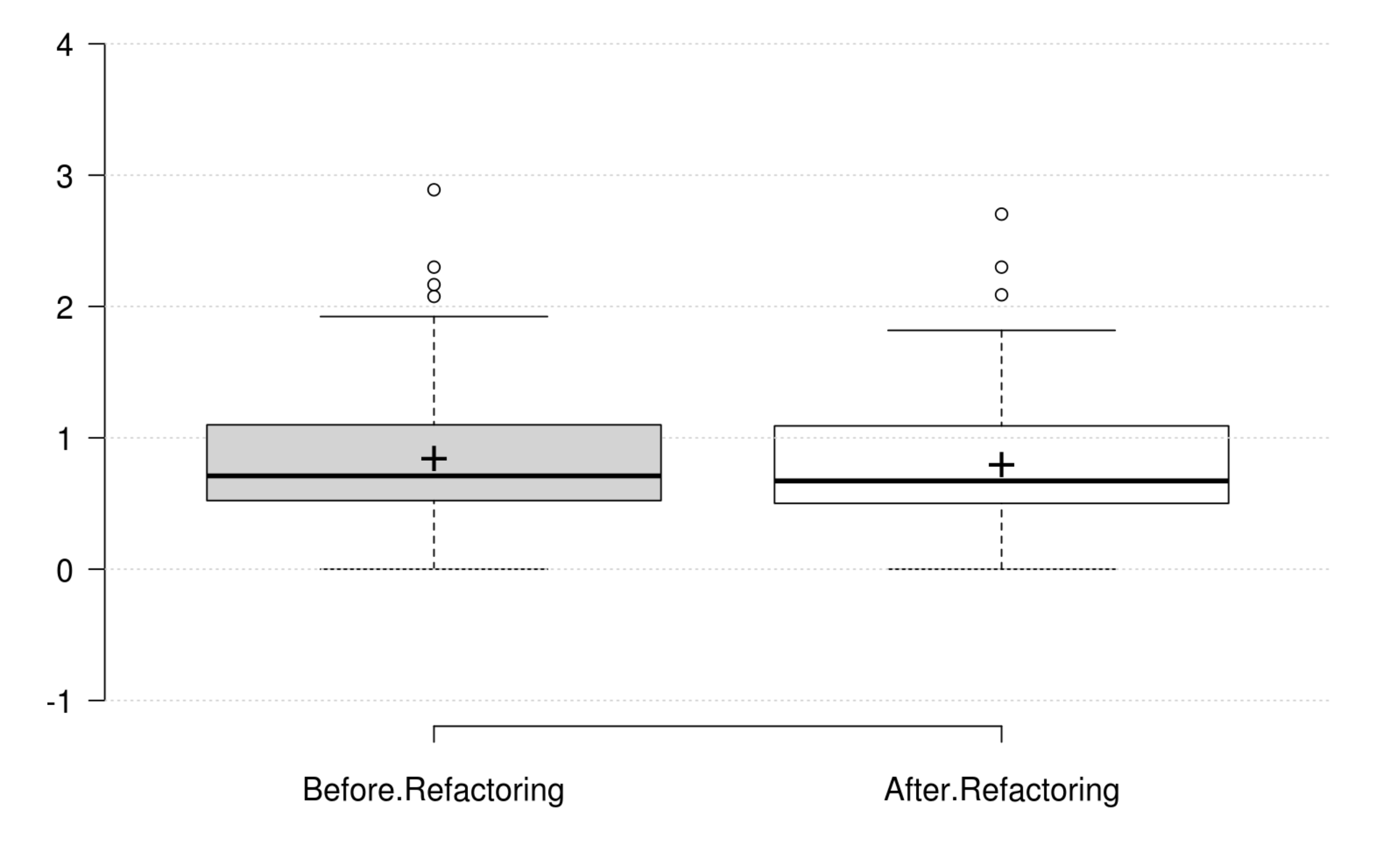}
\caption{Complexity - MaxNest}
\label{BP:complexity-maxnest}
\end{subfigure}%

% row-4
\begin{subfigure}{4.5cm}
\centering\includegraphics[width=4cm]{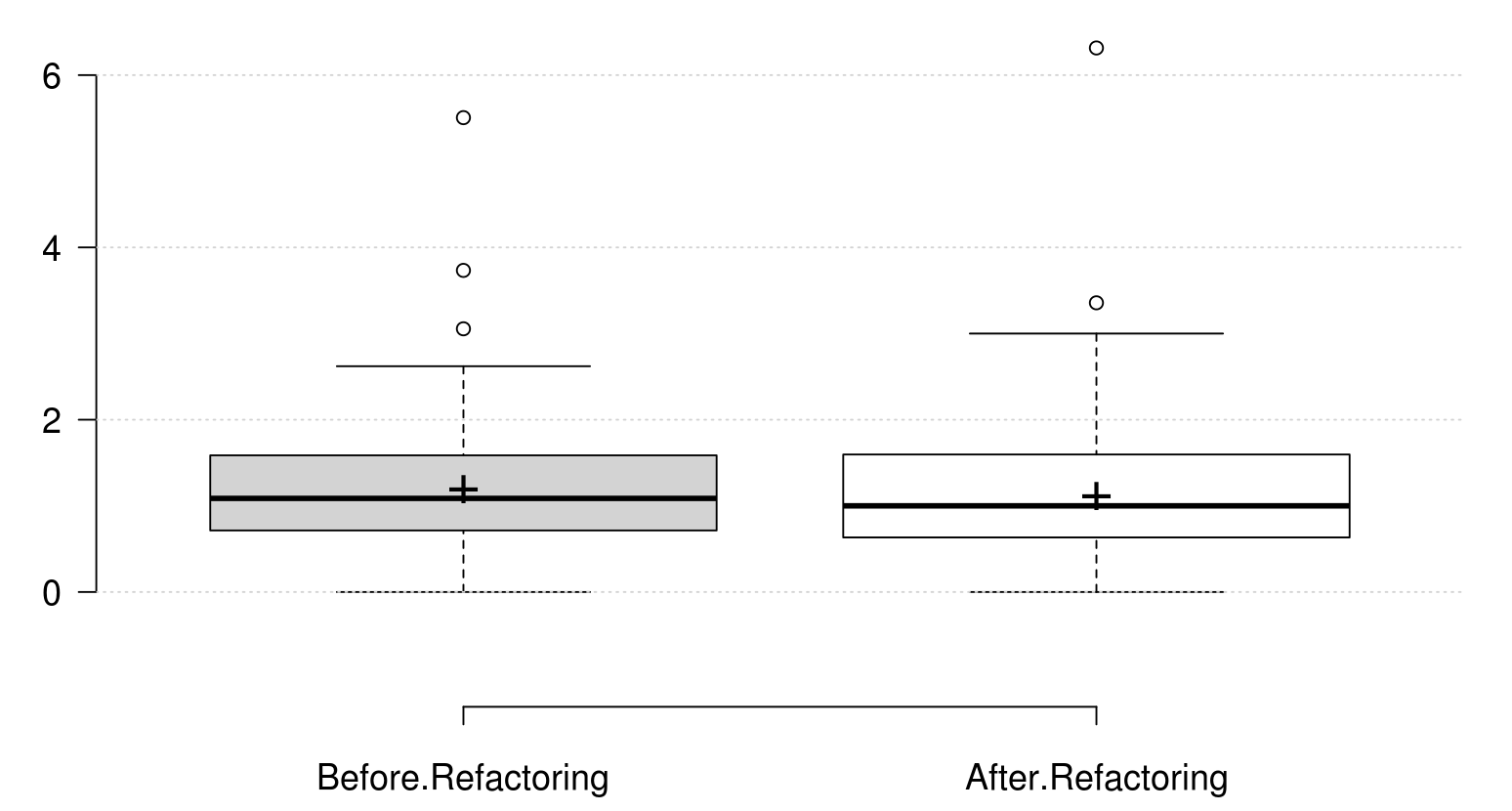}
\caption{Inheritance - DIT}
\label{BP:Inheritance-DIT}
\end{subfigure}
\begin{subfigure}{4.5cm}
\centering\includegraphics[width=4cm]{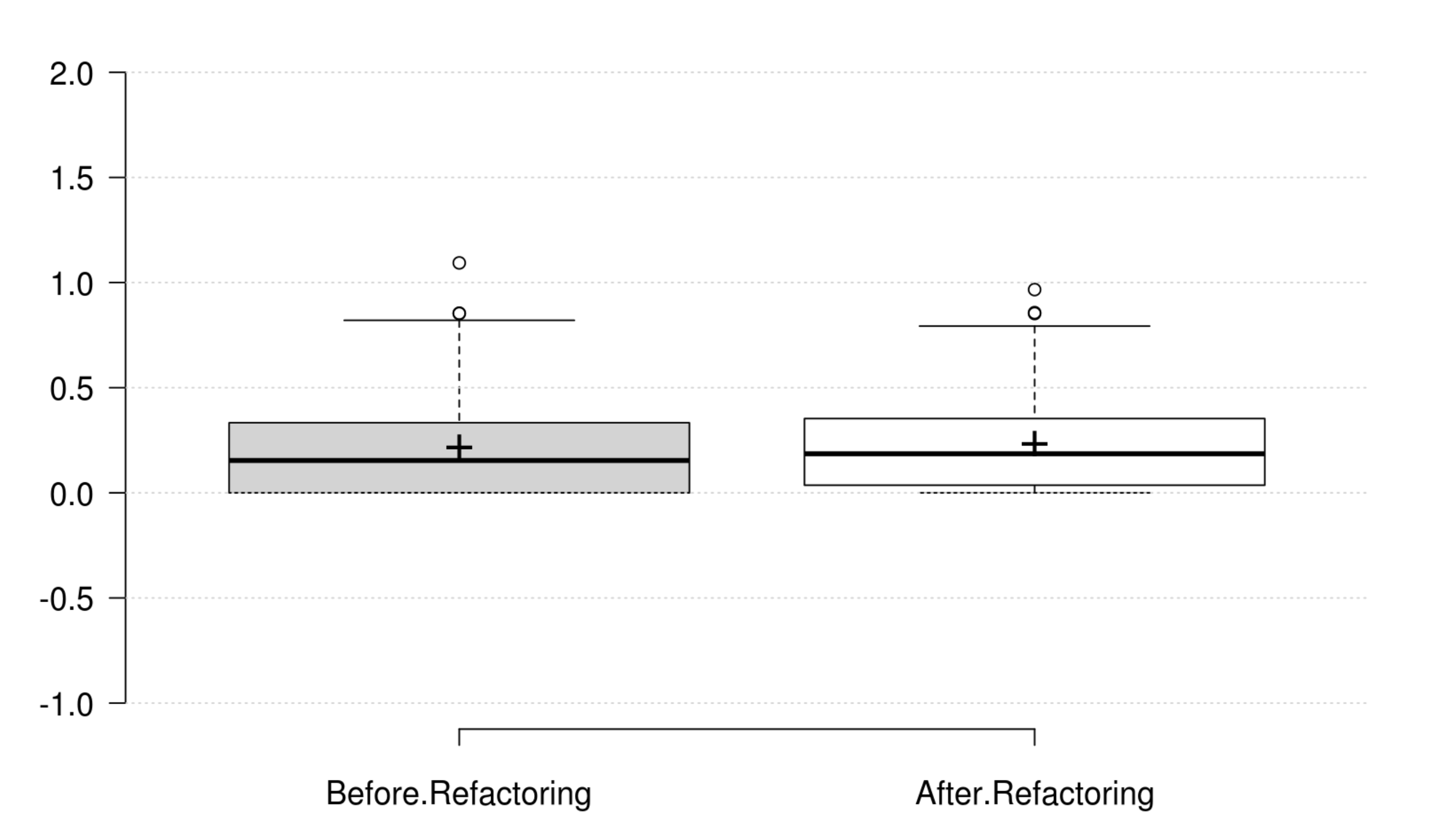}
\caption{Inheritance - NOC}
\label{BP:Inheritance-NOC}
\end{subfigure}%
\begin{subfigure}{4.5cm}
\centering\includegraphics[width=4cm]{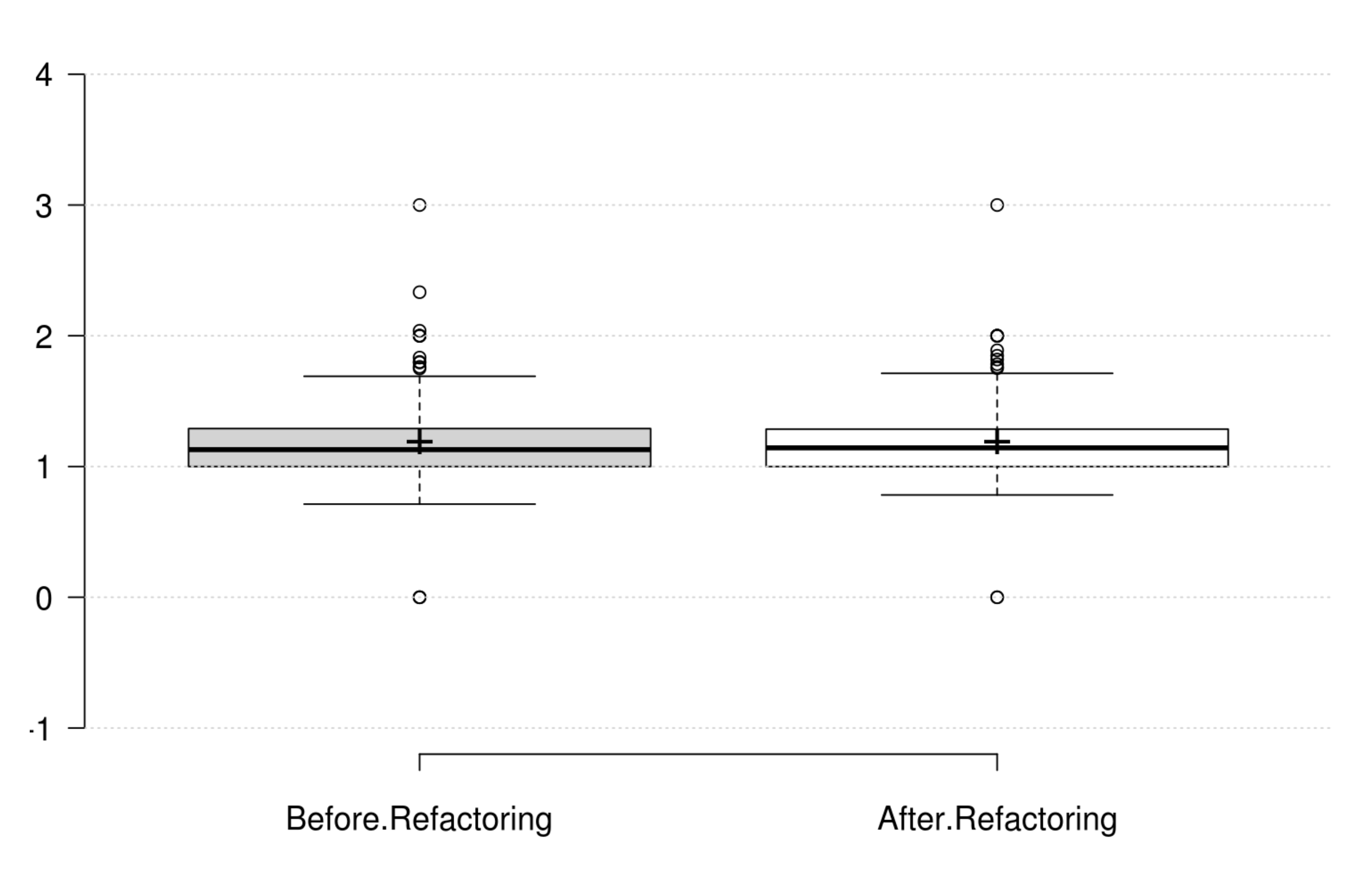}
\caption{Inheritance - IFANIN}
\label{BP:Inheritance-ifanout}
\end{subfigure}%
\begin{subfigure}{4.5cm}
\centering\includegraphics[width=4cm]{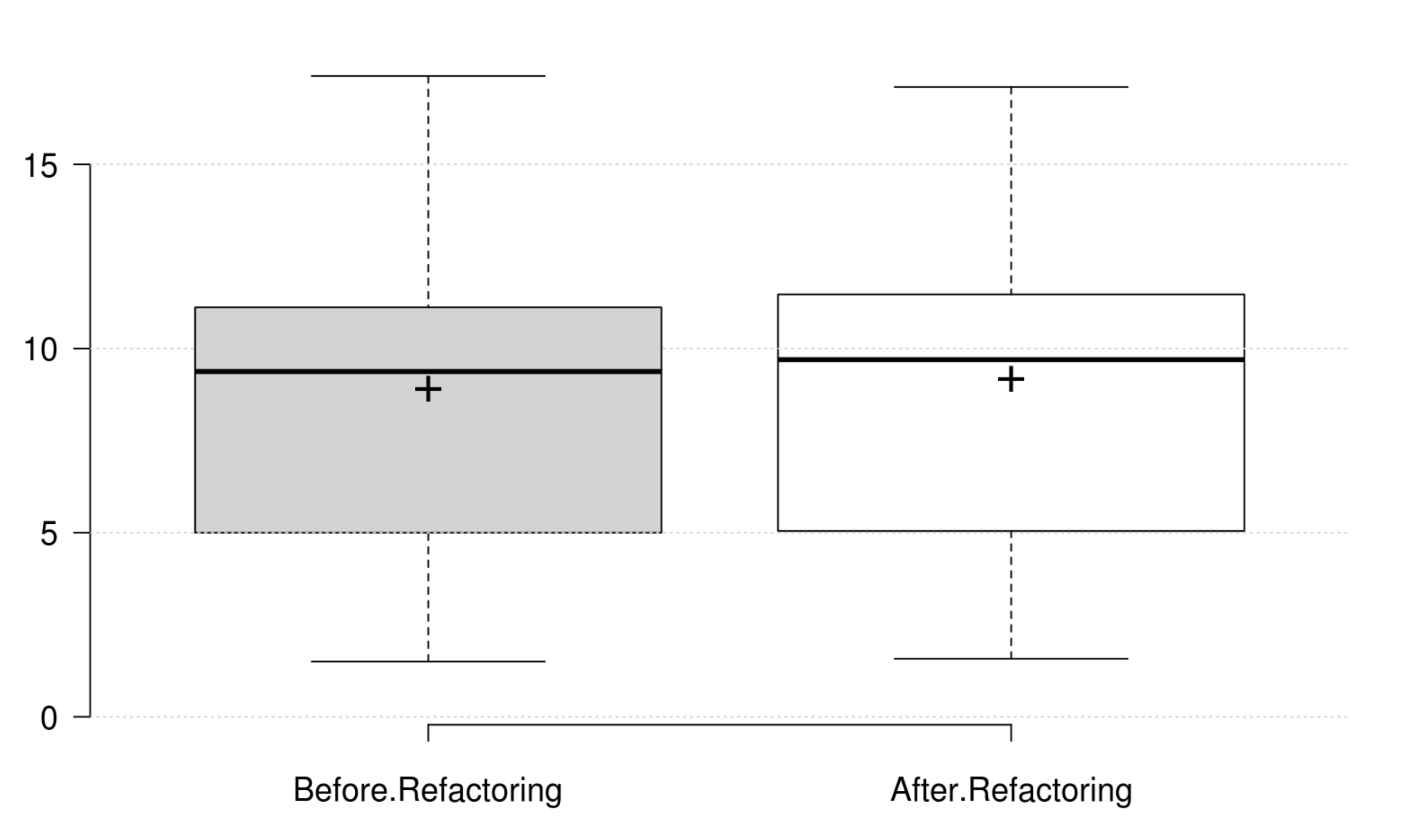}
\caption{Polymorphism - WMC}
\label{BP:polymorphism-WMC}
\end{subfigure}

% row-5
\begin{subfigure}{4.5cm}
\centering\includegraphics[width=4cm]{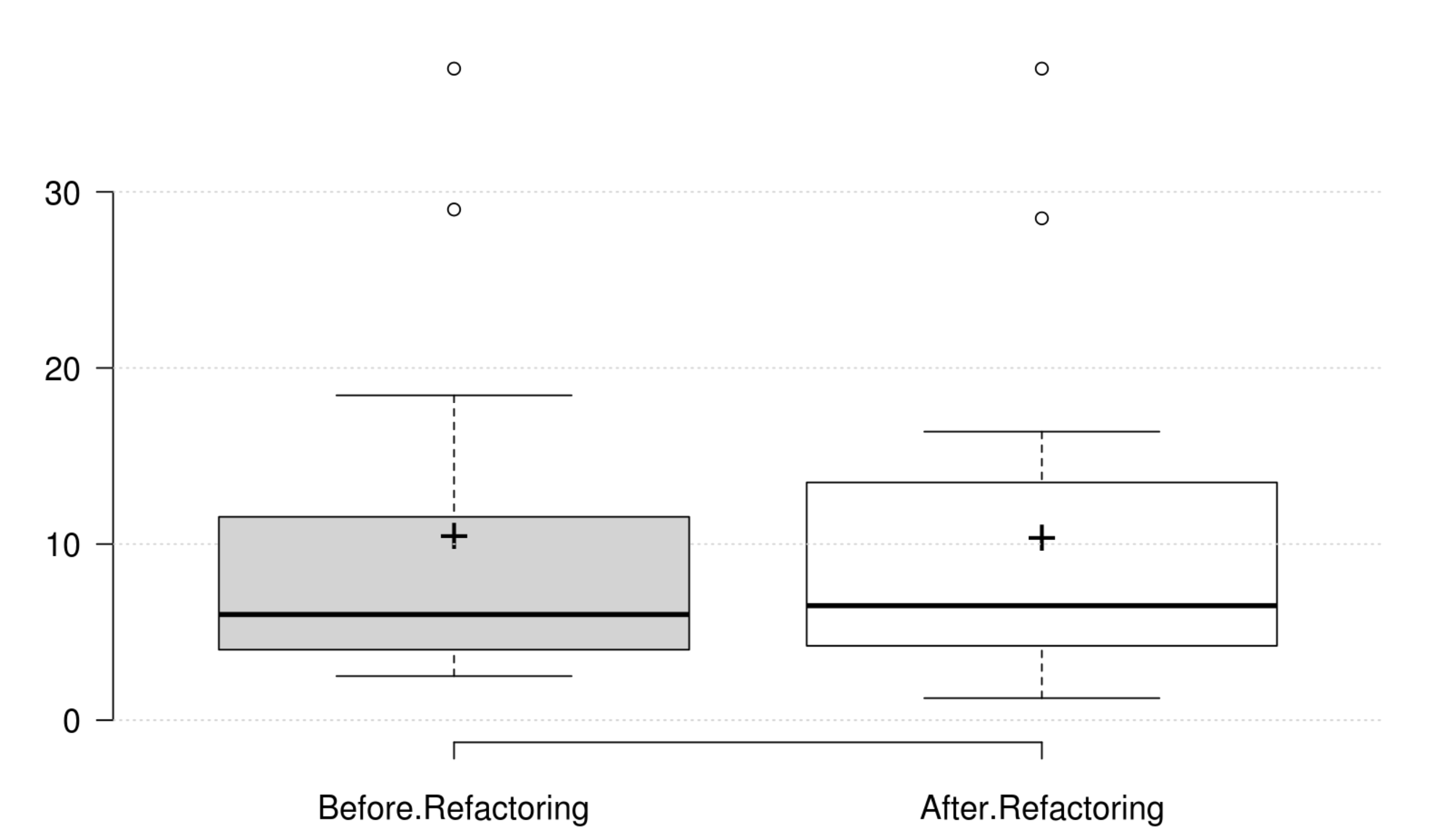}
\caption{Polymorphism - RFC}
\label{BP:polymorphism-rfc}
\end{subfigure}%
\begin{subfigure}{4.5cm}
\centering\includegraphics[width=4cm]{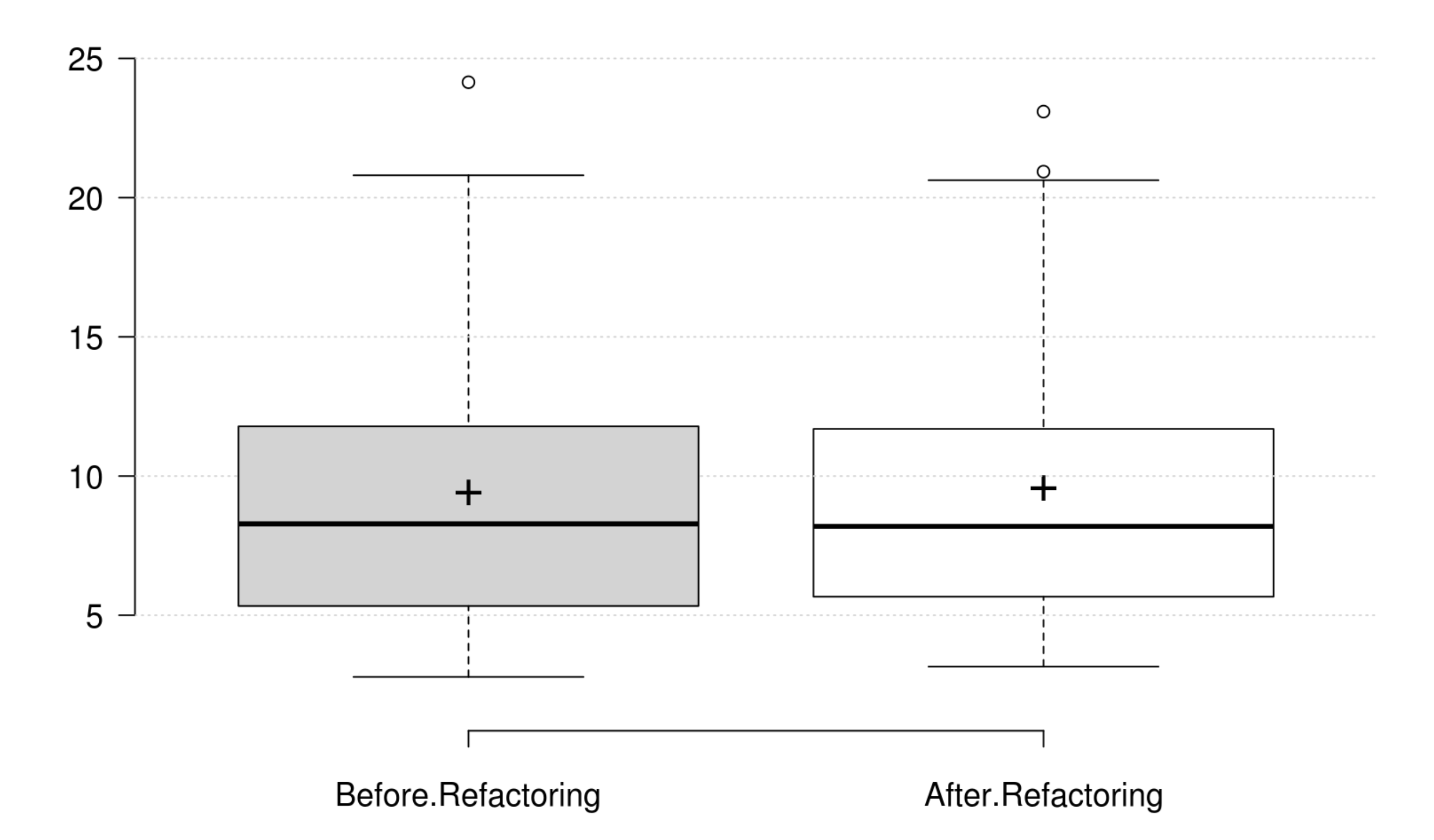}
\caption{Encapsulation - WMC}
\label{BP:Encapsulation-WMC}
\end{subfigure}%
\begin{subfigure}{4.5cm}
\centering\includegraphics[width=4cm]{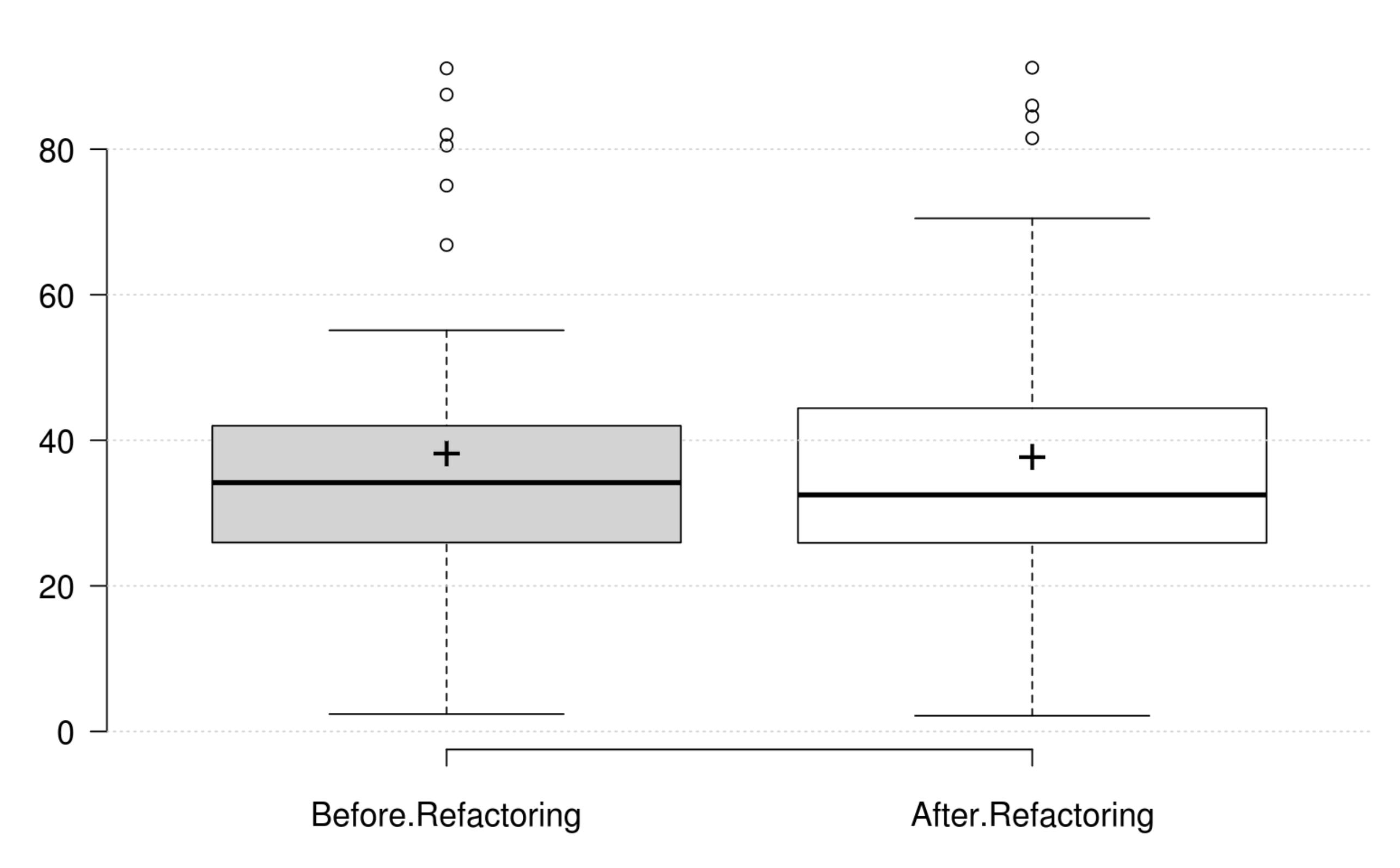}
\caption{Encapsulation - LCOM}
\label{BP:Encapsulation-lcom}
\end{subfigure}
\begin{subfigure}{4.5cm}
\centering\includegraphics[width=4cm]{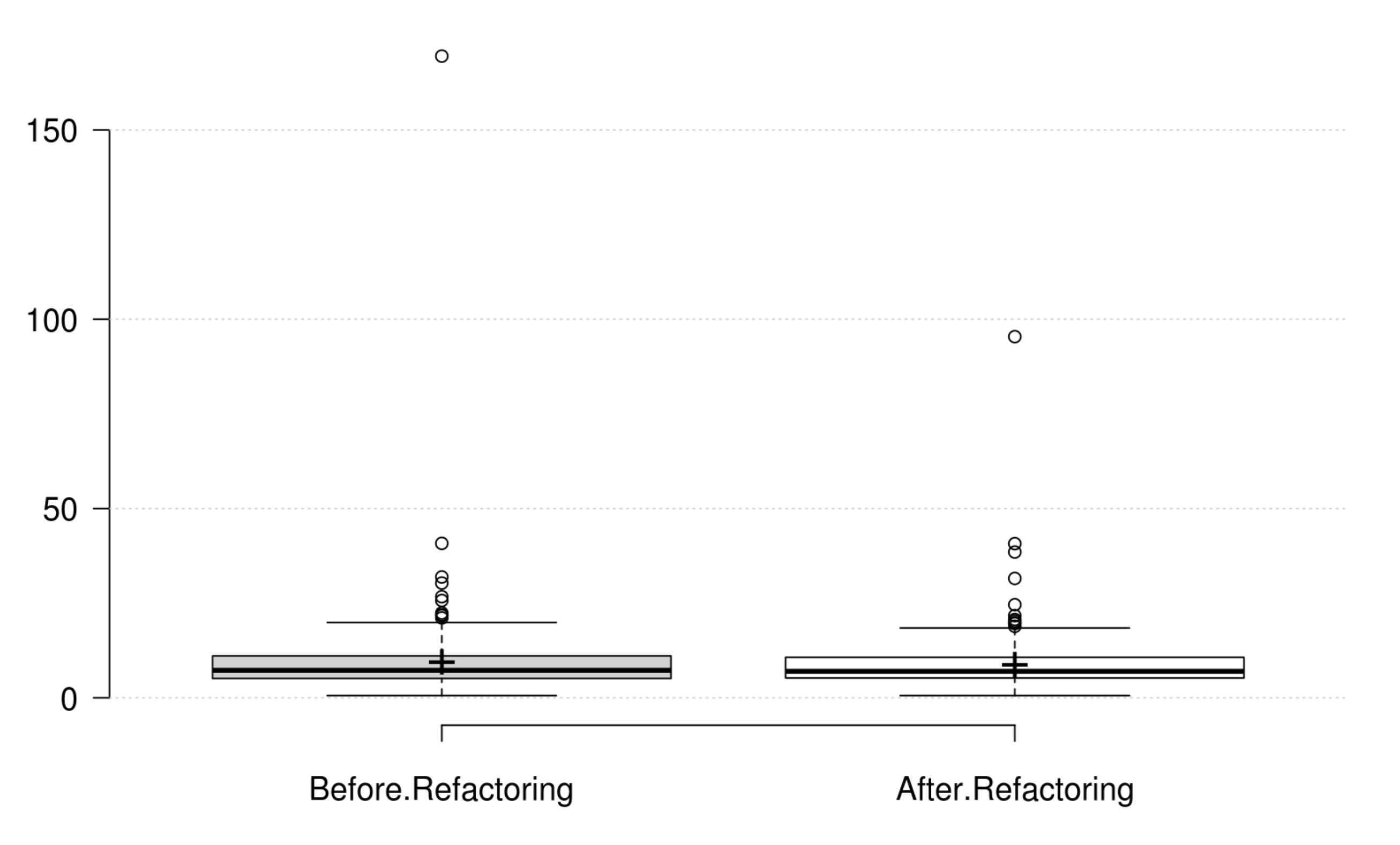}
\caption{Abstraction - WMC}
\label{BP:Abstraction-WMC}
\end{subfigure}%

% row 6
\begin{subfigure}{4.5cm}
\centering\includegraphics[width=4cm]{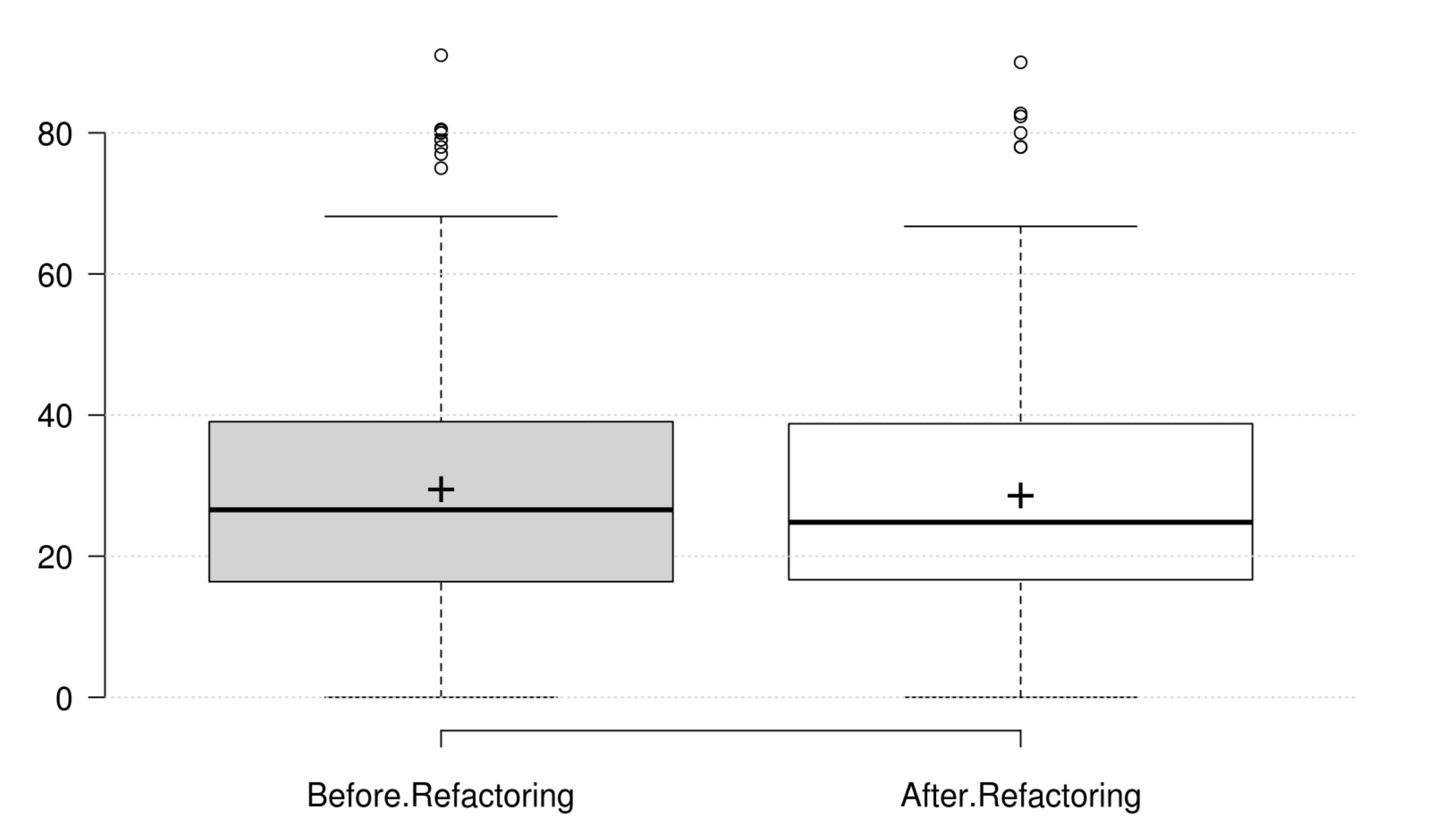}
\caption{Abstraction - LCOM}
\label{BP:Abstraction-lcom}
\end{subfigure}%
\begin{subfigure}{4.5cm}
\centering\includegraphics[width=4cm]{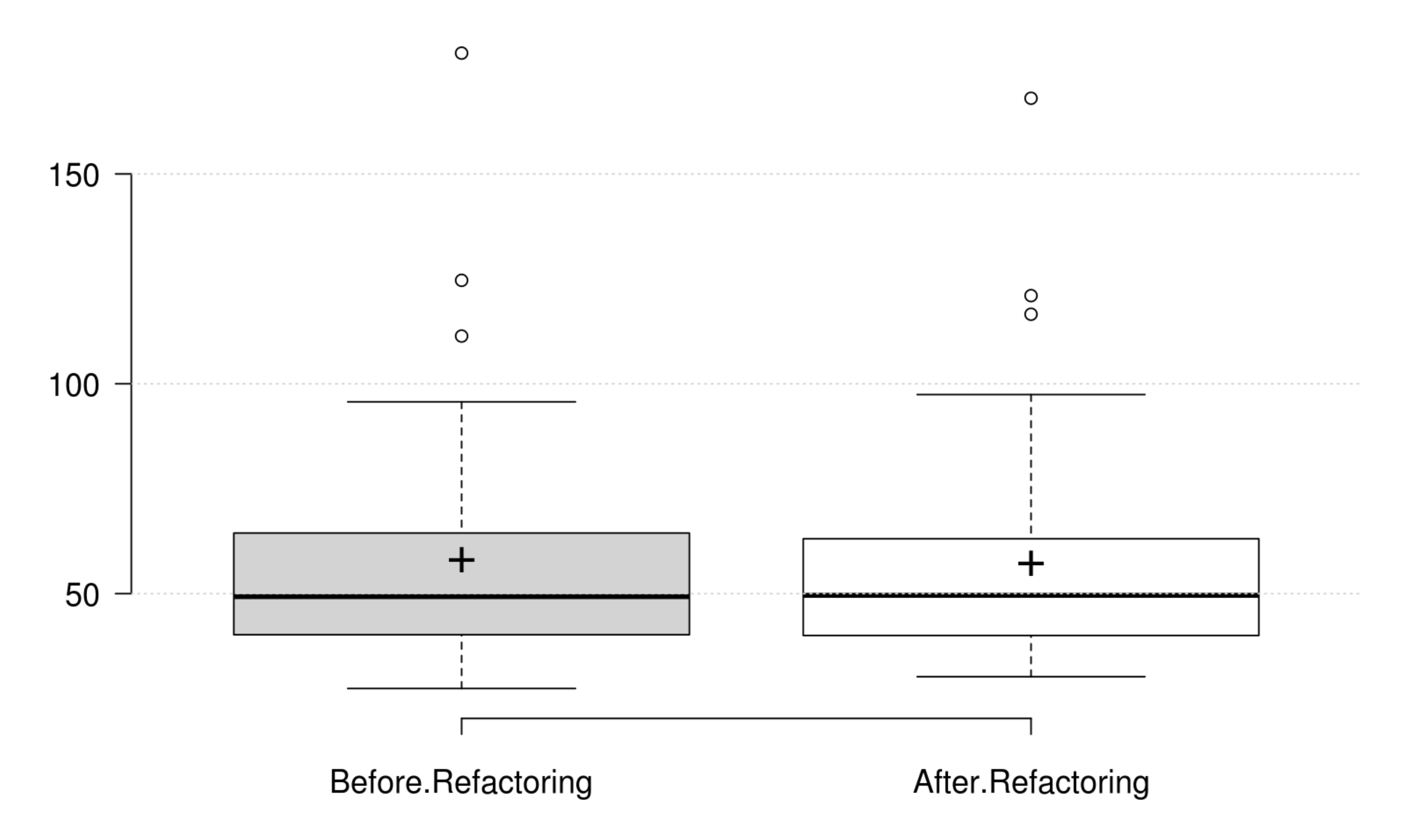}
\caption{Design Size - LOC}
\label{BP:Design Size-loc}
\end{subfigure}
\begin{subfigure}{4.5cm}
\centering\includegraphics[width=4cm]{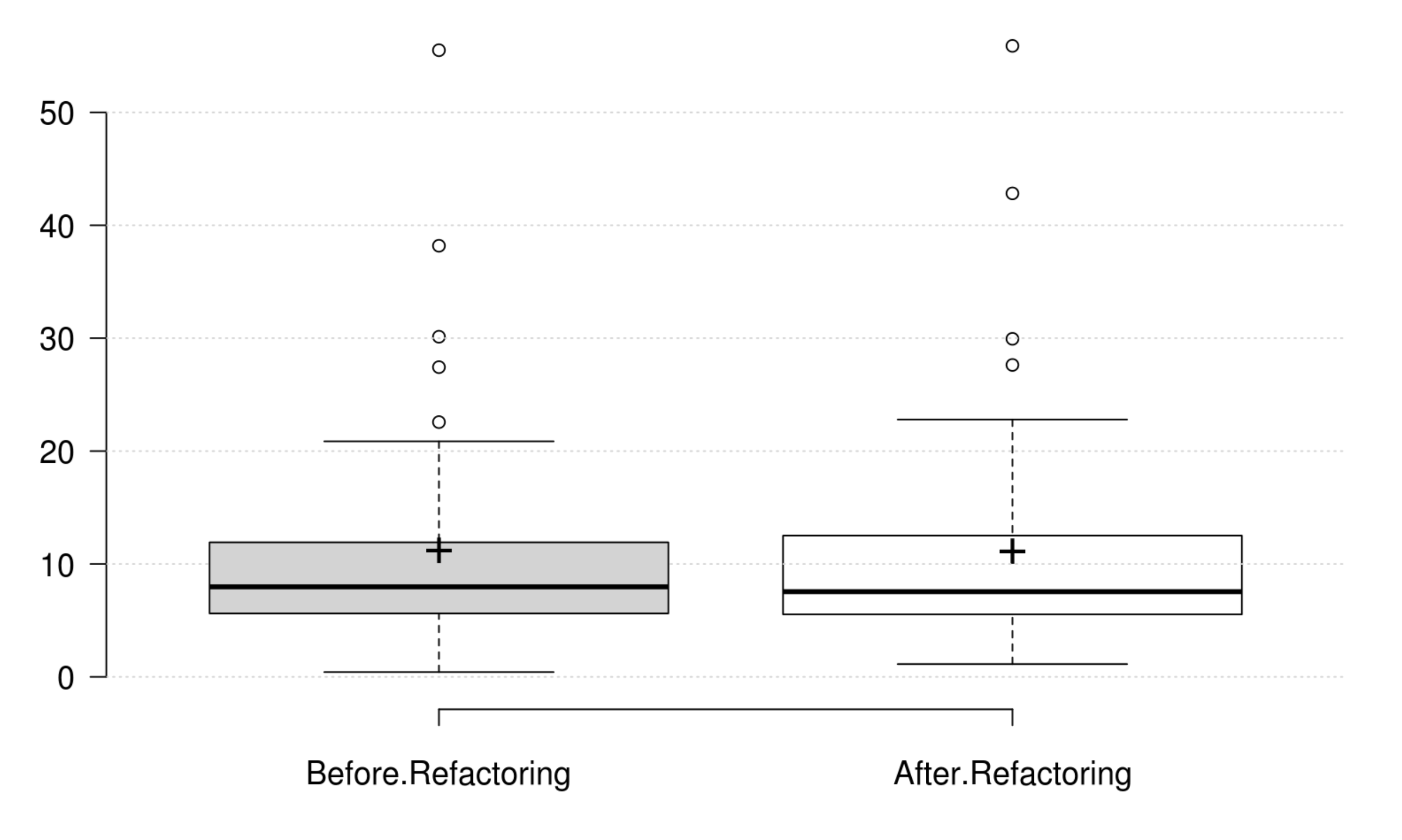}
\caption{Design Size - CLOC}
\label{BP:Design Size-cloc}
\end{subfigure}%
\begin{subfigure}{4.5cm}
\centering\includegraphics[width=4cm]{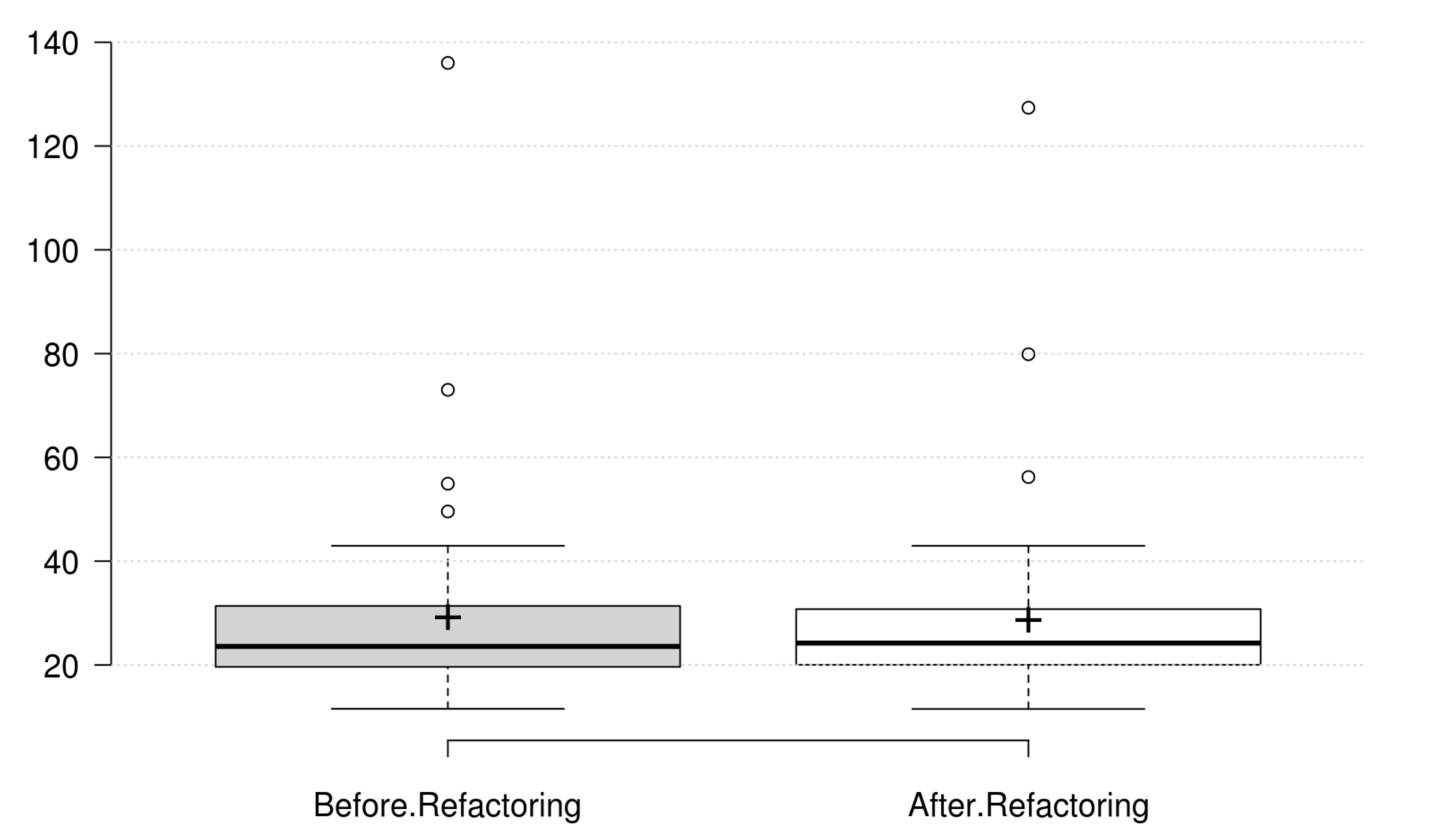}
\caption{Design Size - STMTC}
\label{BP:Design Size-stmtc}
\end{subfigure}%

% row 7
\begin{subfigure}{4.5cm}
\centering\includegraphics[width=4cm]{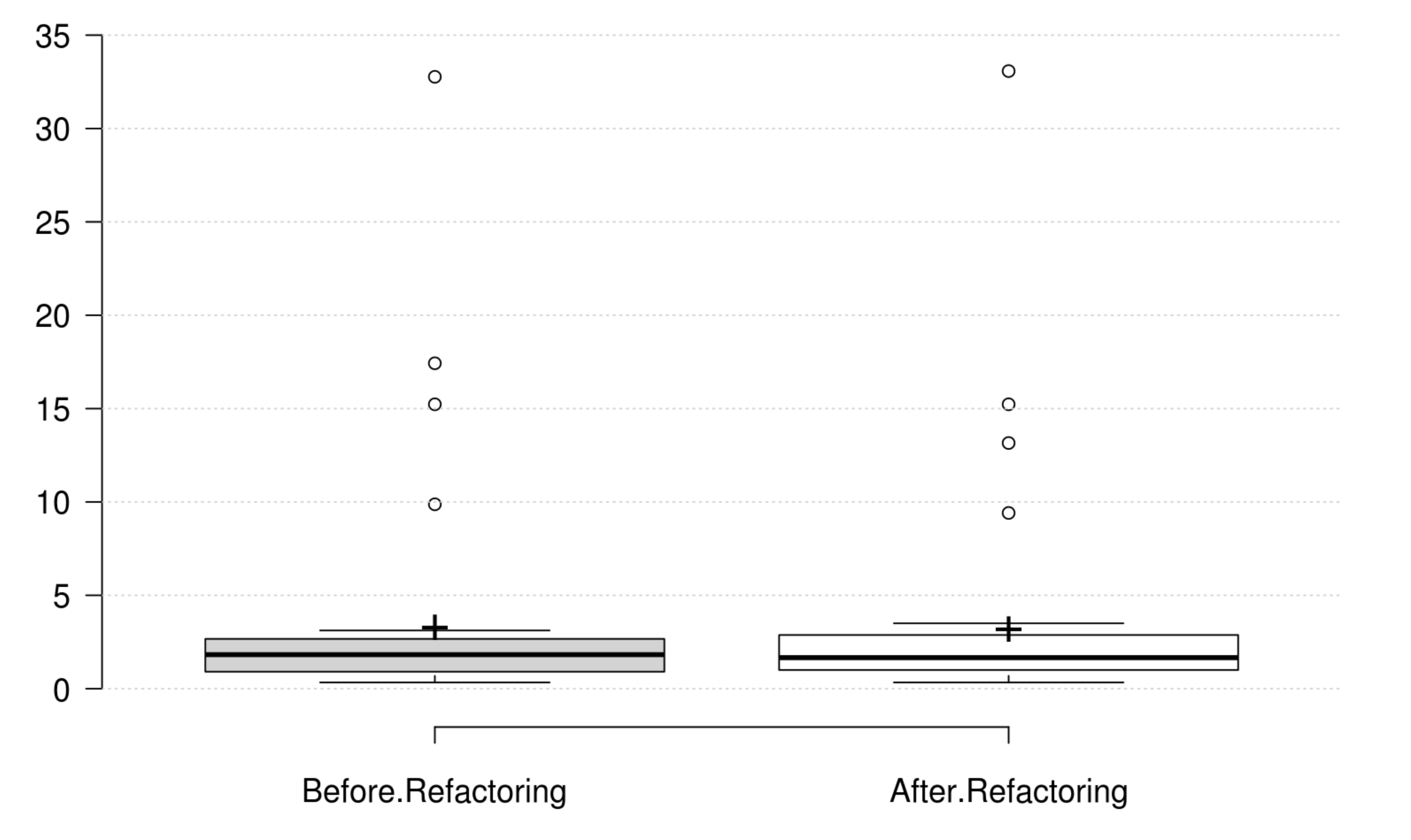}
\caption{Design Size - CDL}
\label{BP:Design Size-cdl}
\end{subfigure}
\begin{subfigure}{4.5cm}
\centering\includegraphics[width=4cm]{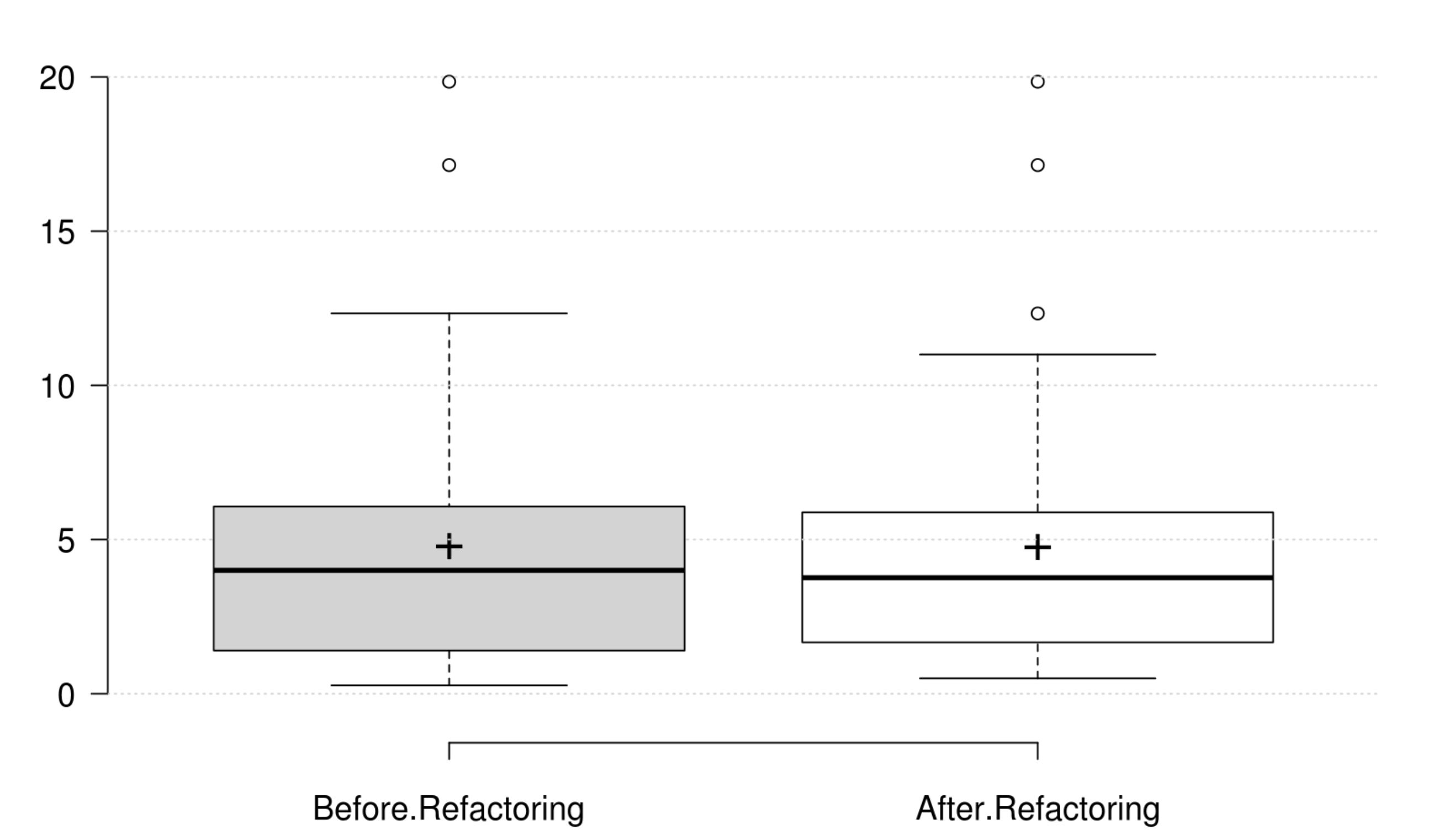}
\caption{Design Size - NIV}
\label{BP:Design Size-niv}
\end{subfigure}%
%\thesubfigure
\begin{subfigure}{4.5cm}
\centering\includegraphics[width=4cm]{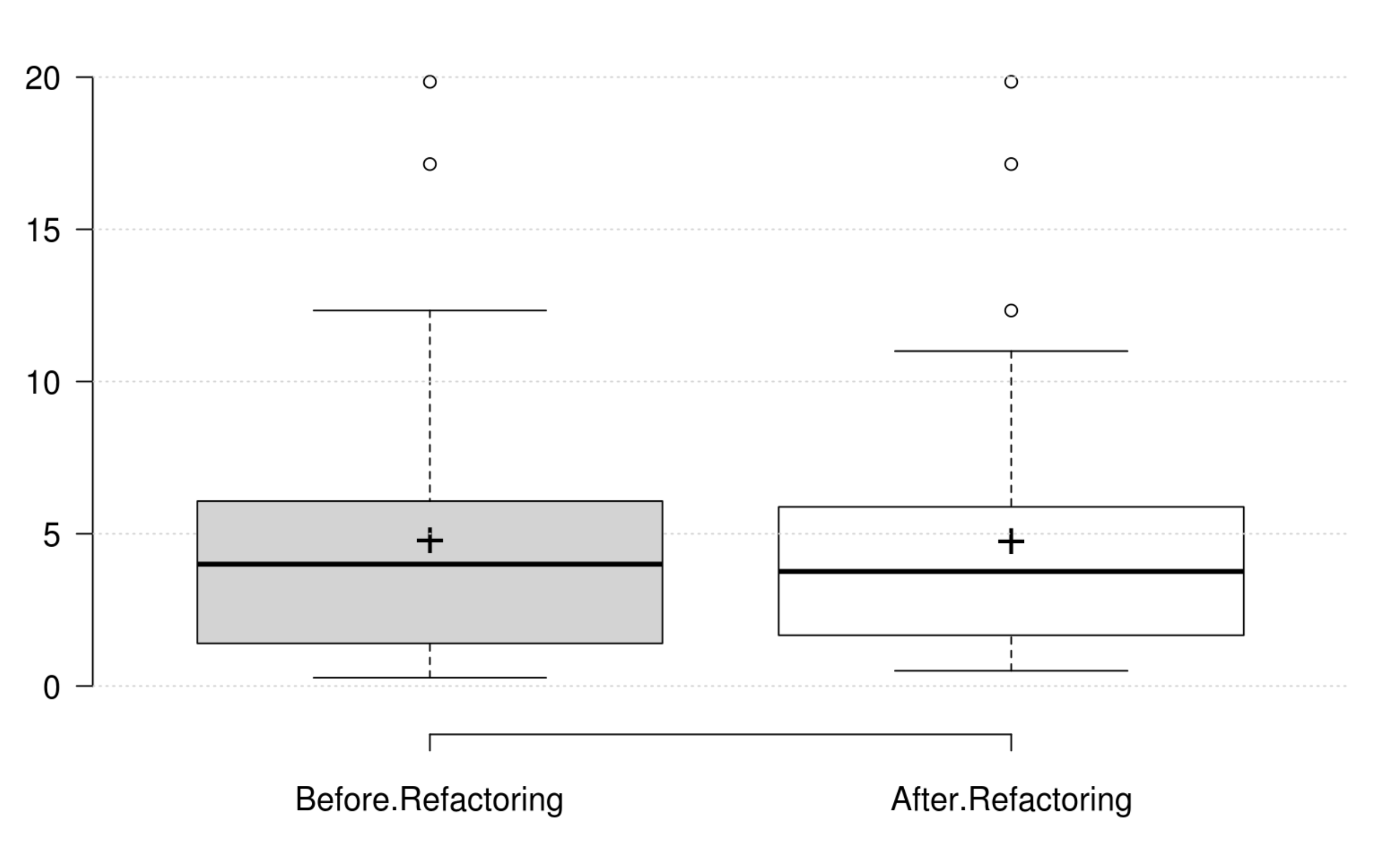}
\caption{Design Size - NIM}
\label{BP:Design Size-nim}
\end{subfigure}%
\caption{Boxplots of metrics values of pre- and post-refactored files.} %\ali{it is good also to add a separate table to show the p-values to easily catch them (and possibly the effect size too)}}\Eman{I actually removed the table to have more space for the discussion. I added it again (Table~\ref{Table:Metrics Suites and Metrics Tools Summary}} 
\label{Chart:Boxplots_Al_V1}
\end{figure*}

% row-8
%\begin{figure}
%\centering
%\begin{subfigure}{4.5cm}
%\centering\includegraphics[width=4cm]{Images/Size-NIM}
%\caption{Design Size - NIM}
%\label{BP:Design Size-nim}
%\end{subfigure}%
%\caption{Metric Values before and after the Commits.} 
%\label{Chart:Boxplots_Al_V2}
%\end{figure}

For each refactoring commit with a documented internal quality attribute by developers, we compute its corresponding metric values (see Table~\ref{Table:Quality Metrics Used in This Study.}) before and after the commit. For instance, for commit messages related to reducing the complexity of the source code, we calculate seven corresponding metric values before and after the selected refactoring commit, \textit{i.e.}, Cyclomatic Complexity (CC), Weighted Method Count (WMC), Response For Class (RFC), Lack of Cohesion of Methods (LCOM), Essential Complexity (Evg), Paths (NPATH), and Nesting (MaxNest) \cite{chidamber1994metrics,mccabe1976complexity,nejmeh1988npath,lorenz1994object}, as shown in Table \ref{Table:Quality Metrics Used in This Study.}. As we calculate the metrics values of pre- and post-refactoring, we want to distinguish, for each metric, whether there is a variation on its pair of values, whether this variation indicates an improvement, and whether that variation is statistically significant. Therefore, we use the Wilcoxon test, a non-parametric test, to compare between the group of metric values before and after the commit, since these groups are dependent on one another. The Null hypothesis is defined by no variation in the metric values of pre- and post-refactored code elements. Thus, the alternative hypothesis indicates that there is a variation in the metric values. In each case, a decreased metric value is considered desirable (i.e., an improvement). Additionally, the variation between values of both sets is considered significant its associated p-value is less than 0.005. It is important to note that, in many cases, the same metric is used to evaluate several quality attributes. %Furthermore, we use the Cliff's Delta (\textit{d}) effect size to estimate the magnitude of the differences. As for its interpretation, we follow the guidelines reported by \cite{trove.nla.gov.au/work/16432558}: negligible for \abs{d} < 0.10, small for 0.10 <= \abs{d} < 0.33, medium for 0.33 <= \abs{d} < 0.474, and large for \abs{d} >= 0.474. 
In the following, we report the results of our research questions.

%\subsection{RQ1: Do the developer perception of quality improvement aligns with the quantitative assessment of code quality?}

The boxplots in Figure~\ref{Chart:Boxplots_Al_V1} show the distribution of each metric before and after each of the examined commits.

To answer our main research question, we provide a detailed analysis of each of the eight quality attributes as reported in Table \ref{Table:Quality Metrics Used in This Study.}. Table~\ref{Table:Metrics Suites and Metrics Tools Summary} shows the overall impact of refactorings on quality.

\subsubsection{Cohesion}
For commits whose messages report the amelioration of the cohesion quality attribute, the boxplot sketched in Figure~\ref{BP:chesion-lcom} shows the pre- and post-refactoring results of the normalized LCOM, used in literature to estimate the cohesion. A poor LCOM metric value implies generally that the classes should be split into 1 or more classes with better cohesion. Thus, if the value of this metric is low, it indicates a strong cohesiveness of the class. We have selected the normalized LCOM metric as it has been widely acknowledged in the literature \cite{pantiuchina2018improving,chavez2017does,henderson1995object} as being the alternative to the original LCOM, by addressing its main limitations (artificial outliers, misperception of getters and setters, etc.). As can be seen from the boxplot in Figure~\ref{BP:chesion-lcom}, the median drops from 28.12 to 25.86 and the third quartile is significantly lower which shows a decrease in variation for commits after refactoring. This result indicates that LCOM is capturing the developer's intention of optimizing the cohesion quality attribute. Furthermore, as shown in Table~\ref{Table:Metrics Suites and Metrics Tools Summary}, LCOM has a positive impact on cohesion quality, as it decreases in the refactored code. This implies that developers did improve the cohesion of their classes, as outlined in their commit messages. %Table~\ref{Table:Metrics Suites and Metrics Tools Summary} shows that the differences in LCOM are statistically significant, but the magnitude of the differences is small.  %However, when we compare the distribution of LCOM before and after the commit messages, the difference is not statistically significant. 
\begin{tcolorbox}
\textit{Summary}. The normalized LCOM metric does not only represent a good replacement to the original LCOM, but also represents the cohesion quality attribute. Its positive variation is in line with the developer's intention in improving cohesion.
\end{tcolorbox}

\subsubsection{Coupling}
For commits whose messages report the amelioration of the coupling quality attribute, the boxplots sketched in Figures~\ref{BP:coupling-cbo},~\ref{BP:coupling-rfc}, ~\ref{BP:coupling-fanin},~\ref{BP:coupling-fanout} show the pre- and post-refactoring results of four structural metrics, \textit{i.e.}, CBO, RFC, FANIN, and FANOUT, used in literature to estimate the coupling. We observe from the figure that three out of the four coupling metrics experienced a degradation in the median values. For instance, CBO, FANIN and FANOUT medians dropped, respectively, from 1.19 to 1.00, from 5.94 to 5.91, and from 2.75 to 2.68. Coupling Between Objects (CBO) counts of the number of classes that are coupled to a particular class either through method or attribute calls. Calls are counted in both directions. CBO values have significantly decreased, which makes it a good representative of coupling. FANIN represents how useful is a code element to other code elements, while FANOUT counts the number of outsider code elements, a particular code element depends on. While both metrics are found to be degrading as developers intend to optimize coupling, only the FANOUT's variation was statistically significant. Interestingly, the Response for a Class (RFC), which counts the visibility of a class to outsider classes, has increased as developers intend to optimize coupling. In theory, increasing the visibility of a class increases the possibility to other classes to reach it, and so, it increases its coupling. However, this does not necessarily hold according to our results, but the variation is not statistically significant.

The manual inspection, of the refactored code, indicates that developers typically decrease coupling by reducing (1) the strength of dependencies that exist between classes, (2) the message flow of the classes, and (3) the number of inputs a method uses plus the number of subprograms that call this method. The code was improved as expected from the developer intentions in their commit message.

\begin{tcolorbox}
\textit{Summary}. CBO, FANIN and FANOUT generally decrease as developer intends to improve coupling. However, only CBO and FANOUT variation is significant. RFC exhibits an opposite variation to coupling, but it is not statistically significant. %Finally, at least one metric has a significant positive variation which matches the developer's perception of improving coupling.
\end{tcolorbox}

%\begin{comment}
%This finding indicate that developers apply refactoring to  decrease coupling which may involve reducing (1) the strength of dependencies that exist between classes, (2) the message flow of the classes, and (3) the number of inputs a method uses plus the number of subprograms that call this method. The code was improved as expected from the developer intentions in their commit message. The differences are statistically significant for CBO and FANIN, but they are not for RFC. The median for FANOUT metric, however, slightly goes up although the difference in the median is not statistically significant. \ali{do you mean not statistically different? if so say it explicitly as you mentioned earlier about Wilcoxon test}. This indicates that developers tend to increase the outputs count which can be in terms of the total number of parameters, global variables, or method calls.     
%\begin{tcolorbox}
%\textit{Summary}. The normalized LCOM metric does not only represent a good replacement to the original LCOM, but also represents the cohesion quality attribute. Its positive variation is in line with the developer's intention in improving cohesion.
%\end{tcolorbox}
%\end{comment}

\subsubsection{Complexity}
As for the complexity quality attribute, we consider seven literature metrics, shown in Table \ref{Table:Quality Metrics Used in This Study.}, to investigate the code complexity reduction as perceived by developers. As seen in the boxplots in Figures ~\ref{BP:complexity-cc},~\ref{BP:complexity-wmc},~\ref{BP:complexity-evg},~\ref{BP:complexity-npath},~\ref{BP:complexity-maxnest}, 
%the complexity metrics values exhibit a reduction after the application of refactoring. Indeed, according to statistically significant positive results for metrics CC, WMC, Evg, NPATH, and MaxNest, the developers tend to reduce complexity. 
%To get a more qualitative sense, we notice from the results both positive and negative trends. 
we observe that the majority metrics \textit{i.e.,} CC, WMC, Evg, NPATH, and MaxNest, experienced a degradation in the median values. Furthermore, all the variations are statistically significant. Despite being associated with several, metrics, which are different in their definitions, our results indicate that 5 out the 7 metrics, accurately represent the complexity quality attribute. However, RFC's opposed increase is found to be statistically significant. %For instance, Cyclomatic Complexity (CC) is one of the most popular complexity measurements, introduced by Thomas McCabe \cite{mccabe1976complexity} which counts the number of predicates (binary decision points) in the program plus one.

In particular, through a manual inspection of the collected dataset, we observe that developers tend to reduce the number of local methods, simplify the structure statements, reduce the number of paths in the body of the code, and lower the nesting level of the control statements (\textit{e.g.}, selection and loop statements) in the method body. On the other hand, when we observe a significant increase in RFC, we notice that developers lower the complexity of methods by pulling them up in the hierarchy, and so they increase the number of inherited methods. %\ali{this last sentence is not very clear}.
\begin{tcolorbox}
\textit{Summary}. CC, WMC, Evg, NPATH, and MaxNest generally decrease as developer intends to improve complexity, and all their variation is significant. Furthermore, our empirical investigation discards RFC from being an indicator for complexity. %Finally, at least one metric has a significant positive variation which matches the developer's perception of improving complexity.
\end{tcolorbox}

\subsubsection{Inheritance}

For commits with amelioration to the inheritance quality attribute, the boxplots sketched in Figures~\ref{BP:Inheritance-DIT},~\ref{BP:Inheritance-NOC},~\ref{BP:Inheritance-ifanout} show the pre- and post-refactoring results of three structural metrics, \textit{i.e.}, DIT, NOC, and IFANIN, used in literature to estimate the inheritance. We observe that only one metric out of the three experienced a degradation in the median values. For instance, the median decreases from 1.09 to 1.00 for DIT, whereas the medians increase from 0.15 to 0.19 and from 1.13 to 1.14 for NOC and IFANIN respectively. This indicates that developers probably decrease the depth of the hierarchy by adding more methods for a class to inherit, increasing the number of immediate subclasses, and increasing the number of immediate base classes. Although we observed certain cases that show inheritance improvement as perceived by developers, the overall depth of the inheritance tree and the number of immediate subclasses and superclasses did not decrease. The interpretation of the metric improvement highly depends on the quality of the code and the developer's design decisions. The statistical test shows that the differences are statistically significant for DIT and NOC, but they are not for IFANIN.

\begin{tcolorbox}
\textit{Summary}. DIT generally decreases as developer intends to improve inheritance, and its variation is significant. IFANIN exhibit opposite variations to inheritance, but it is not statistically significant. Furthermore, our empirical investigation discards NOC from being an indicator for inheritance. %Finally, at least one metric has a significant positive variation which matches the developer's perception of improving inheritance.
\end{tcolorbox}

%For inheritance, we tested the commits that explicitly refer to the inheritance improvement by either adding or removing it using three metrics (\textit{i.e.}, DIT, NOC, and IFANIN). As shown in the box plots (~\ref{BP:Inheritance-DIT},~\ref{BP:Inheritance-NOC},~\ref{BP:Inheritance-ifanout}), the median decreases from 1.09 to 1.00 for DIT, whereas  the medians increase from 0.15 to 0.19 and from 1.13 to 1.14 for NOC and IFANIN respectively. This indicates that developers probably decrease the depth of the hierarchy by adding more methods for a class to inherit, increasing the number of immediate subclasses, and increasing the number of immediate base classes. Although we observed certain cases that show inheritance improvement as perceived by developers, the overall depth of the inheritance tree and the number of immediate subclasses and superclasses did not decrease. The interpretation of the metric improvement highly depends on the the quality of the code and the developer's design decisions. The statistical test shows that the differences are statistically significant for DIT and NOC, but they are not for IFANIN.

\subsubsection{Polymorphism}

For commits whose messages report the amelioration of the polymorphism quality attribute, the boxplots sketched in Figures~\ref{BP:polymorphism-WMC},~\ref{BP:polymorphism-rfc} show the pre- and post-refactoring results of two structural metrics, \textit{i.e.}, WMC and RFC, used in literature to estimate the polymorphism. We observe that none of these metrics experienced a degradation in the median values. 

The concept of polymorphism is closely related to inheritance. When developers inherit instance variables and methods from another class, polymorphism techniques allow the subclasses to use these variables and methods to perform different tasks. For this quality attribute, we observe similar trends to inheritance. There is a rise in the median for both WMC and RFC. When developers explicitly refer to polymorphism aspect improvement as a target in the commit messages, they tend to increase the number of local and inherited methods. The statistical test shows that the differences are not statistically significant. 

\begin{tcolorbox}
\textit{Summary}. WMC and RFC exhibit opposite variations to polymorphism, but they are not statistically significant. Therefore, we could not find any metric that has a significant positive variation which matches the developer's perception of improving polymorphism.
\end{tcolorbox}

\subsubsection{Encapsulation}

For commits whose messages report the amelioration of the encapsulation quality attribute, the boxplots sketched in Figures~\ref{BP:Encapsulation-WMC},~\ref{BP:Encapsulation-lcom} show the pre- and post-refactoring results of two structural metrics, \textit{i.e.}, WMC and the normalized LCOM, used in literature to estimate the encapsulation. We observe that both metrics experienced a degradation in the median values. However, the variations are statistically significant.

%Encapsulation is the quality attribute for which we observed an improvement for both of the selected metrics (i.e., WMC and LCOM) with a drop of its WMC and its LCOM median. This indicates that developers prevent access to attributes and methods by defining them to be private and enclosing them within a single construct. Although the results for the encapsulation metrics are not statistically significant, the significant results for cohesion and complexity-related commits discussed previously might indicate that the information hiding mechanism could generally help in reducing the complexity of the software systems when developers are actually limiting the inter-dependencies between components, and thus promote cohesion and modularity.   %\Eman{need to link encapsulation to cohesion since we use LCOM to test the improvement}. 

From a qualitative perspective, we observe that developers prevent access to attributes and methods by defining them to be private and enclosing them within a single construct. Although the results of the encapsulation metrics are not statistically significant, the significant results of cohesion and complexity-related commits discussed previously might indicate that the information hiding mechanism could generally help in reducing the complexity of the software systems when developers are actually limiting the inter-dependencies between components, and thus promote cohesion and modularity.

\begin{tcolorbox}
\textit{Summary}. WMC and the normalized LCOM generally decrease as developer intends to improve encapsulation, but their variations are not significant. Therefore, we could not find any metric that has a significant positive variation which matches the developer's perception of improving encapsulation.
\end{tcolorbox}

\subsubsection{Abstraction}

For this quality attribute that measures the generalization-specialization aspect of the design, we noticed an improvement of both the WMC and the normalized LCOM metrics, as shown in Figures~\ref{BP:Abstraction-WMC},~\ref{BP:Abstraction-lcom}. The differences are not statistically significant. Using this attribute, developers seem to practically handle the complexity of the methods when adding one or more descendants by actually hiding the implementation details, and increasing the class cohesion.
%\Eman{not really sure if more or less abstraction is good or bad changes}. 

\begin{tcolorbox}
\textit{Summary}. WMC and the normalized LCOM generally decrease as developer intends to improve abstraction, but their variations are not significant. Therefore, we could not find any metric that has a significant positive variation which matches the developer's perception of improving abstraction.
\end{tcolorbox}

\subsubsection{Design Size}

For commits whose messages report the amelioration of the design size quality attribute, the boxplots sketched in Figures~\ref{BP:Design Size-loc},~\ref{BP:Design Size-cloc},~\ref{BP:Design Size-stmtc},~\ref{BP:Design Size-cdl},~\ref{BP:Design Size-niv},~\ref{BP:Design Size-nim} show the pre- and post-refactoring results of five structural metrics, \textit{i.e.}, LOC, CLOC, STMTC, CDL, NIV, NIM, used in literature to estimate the design size. We notice the improvement of four metrics, namely CLOC, CDL, NIV, and NIM after the commits in which developers explicitly target the improvement of the size of the classes. As can be seen in the box plots, the medians decreased in general. On the other hand, we notice an increase in LOC and STMTC. Regardless of the increase or decrease of metric values, their variations are not statistically significant. This indicates that developers reduce (1) line containing comments, (2) the number of classes and (3) the  number of declared instance variables and methods. As for LOC and STMTC, we observed minor increases in the metric values. %This shows us that developers added more lines of code plus more declarative and executable statements after the application of refactoring that might be because developer intentions is to improve the readability and the clarity of the code. 

\begin{tcolorbox}
\textit{Summary}. CLOC, CDL, NIV, and NIM generally decrease as developers intend to improve design size, but their variations are not significant. Therefore, we could not find any metric that has a significant positive variation which matches the developer's perception of improving design size.
\end{tcolorbox}

% boxplots used to be here.

%\input{Tables/QualityImpactSummary.tex}

\begin{table}
  \centering
\caption{Effect of refactoring on structural metrics, clustered by their corresponding internal quality attribute. (+ve) indicates positive impact; (+ve) indicates negative impact; \textbf{bold} indicates statistical significance; \textit{italic} indicates improvement.}
\label{Table:Metrics Suites and Metrics Tools Summary}
%\begin{sideways}
%\begin{adjustbox}{width=1.1\textwidth,center}
%\begin{adjustbox}{width=\textheight,totalheight=\textwidth,keepaspectratio}
\begin{tabular}{llll}\hline
\toprule
\bfseries Quality Attribute & \bfseries Metric & \bfseries Impact & \bfseries \textit{p}-value \\
\midrule
%\multicolumn{2}{l}{\textbf{\textit{Internal Quality Attribute }}}\\
%\midrule
Cohesion & LCOM  & +ve & \textit{\textbf{0.0346}} %& 0.2244898
%(Small)   
\\ 
Coupling & CBO  & +ve & \textit{\textbf{0.0400}} %& 0.1010702
%(Small) 
\\
         & RFC & -ve & 0.2729 %& -0.008034026
%(Negligible) 
\\
         & FANIN & +ve & \textit{0.2338} 
%& 0.03579868
%(Negligible) 
\\     
         & FANOUT & +ve & \textbf{\textit{0.0456}} %& 
%0.006143667
%(Negligible) 
\\
Complexity & CC & +ve & \textit{\textbf{0.0001}} %& 0.06425956
%(Negligible) 
\\
           & WMC & +ve & \textit{\textbf{0.0062}} %& -0.03332913
%(Negligible) 
\\
           & RFC & -ve &  \textbf{0.0021} 
%& -0.04532256
%(Negligible) 
\\
           & LCOM & -ve & 0.2431  
%& -0.04003099 (Negligible) 
\\
           & Evg & +ve &  \textit{\textbf{0.0010}} 
%&0.03774776
%(Negligible) 
\\
           & NPATH & +ve & \textit{\textbf{< 0.0001}} 
%&0.1031435
%(Small) 
\\
           & MaxNest & +ve &  \textit{\textbf{0.0026}}  
%&0.04746875
%(Negligible) 
\\
Inheritance & DIT & +ve & \textit{\textbf{0.0439}}  
%& 0.09559221
%(Negligible) 
\\
            & NOC & -ve & \textbf{0.0208}  
%& -0.05786504
%(Negligible) 
\\
            & IFANIN & -ve & 0.3987 
%& -0.005076256
%(Negligible) 
\\ 
Polymorphism & WMC & -ve & 0.5137 
%& -0.05190311
%(Negligible) 
\\
             & RFC & -ve & 0.7983 
%& -0.03114187
%(Negligible)
\\
Encapsulation & WMC & +ve & \textit{0.1769} 
%& -0.04705215
%(Negligible) 
\\
              & LCOM & +ve & \textit{0.7737} 
%& 0.04195011
%(Negligible) 
\\
Abstraction & WMC & +ve & \textit{0.1924} 
%& 0.030248
%(Negligible) 
\\
            & LCOM & +ve & \textit{0.6988} 
%& 0.02403461
%(Negligible) 
\\ 
Design Size & LOC & -ve &  0.8245 
%& 0.01473923
%(Negligible) 
\\
            & CLOC & +ve & \textit{0.7855} 
%& 0.04365079
%(Negligible)  
\\
            & STMTC & -ve & 0.3311  
%& -0.01870748
%(Negligible)
\\
            & CDL & +ve & \textit{0.4870}
%& 0.000566893
%(Negligible)
\\
            & NIV & +ve & \textit{0.2757} 
%& 0.002834467
%(Negligible)
\\
            & NIM & +ve & \textit{0.6043} 
%& -0.01530612
%(Negligible)
\\
\bottomrule
\end{tabular}
%\end{adjustbox}
%\end{sideways}
\end{table}

\section{Threats to Validity}
\label{sec:Threats}
%We identify, in this section, potential threats to the validity of our approach and our experiments. 
Our study has used a few thousands of refactoring commits in various systems. Since the analysis was not carried out in a controlled environment, few threats are discussed in this section as follows: 

\textbf{Internal Validity.} Our analysis is mainly threatened by the accuracy of the refactoring mining tools because the tool may miss the detection of some refactorings. However, previous studies \cite{silva2016we,tsantalis2018accurate} report that Refactoring Miner has high precision and recall scores compared to other state-of-the-art refactoring detection tools, which gives us confidence in using the tool. Another potential threat to validity relates to commit messages. This study does not exclude commits containing tangle code changes \cite{herzig2016impact}, in which developers performed changes related to different tasks and one of these tasks could be related to quality enhancement. If these changes were committed at once, there is a possibility that the individual changes are merged and cannot trace it back to the original task. We did not consider filtering out such changes in this study. Moreover, our manual analysis is a time consuming and error prone, which we tried to mitigate by focusing mainly on commits known to contain refactorings. 

Another potential threat to validity is the sample bias, where the choice of the data may directly impact the results. Therefore, we explored a large sample of projects, we made sure they are well engineered to ensure the quality of the findings along with diversifying the sources to reduce the bias of data belonging to the same entity. During our qualitative analysis, we considered only commits where a consensus between authors was made about whether a message is clearly stating the enhancement of a particular quality attribute. Commits which were debatable were discarded. We also provide our dataset online for further refinement and analysis.

\textbf{Construct Validity.} A potential threat to construct validity relates to the set of metrics, as it may miss some properties of the selected internal quality attributes. To mitigate this threat, we select well-known metrics that cover various properties of each attribute, as reported in the literature \cite{chidamber1994metrics}.

\textbf{External Validity.} Our analysis was limited to only open-source Java projects. However, we were still able to analyze 3,795 projects that are well-commented, and varied in size, contributors, number of commits and refactorings.

\section{Conclusion}
\label{sec:conclusion}
%Software developers do explicitly report target quality improvements in the commit messages of versioned repositories. 
In this work, we performed an exploratory study to investigate the alignment between quality improvement and software design metrics by focusing on 8 internal quality attributes and 27 structural metrics. In summary, the main conclusions are:
%\begin{itemize}
    %\item A variety of structural metrics can represent the internal quality attributes considered in this study. Based on our empirical investigation, for metrics that are associated with quality attributes, there are different degrees of improvement and degradation of software quality. For instance.
    %\item Most of the metrics that are mapped to the main quality attributes, \textit{i.e.,} cohesion, coupling, and complexity, do capture developer intentions of quality improvement reported in the commit messages. In contrast, there is also a case in which the metrics do not capture quality improvement as perceived by developers. We summarize our findings as follows: %demonstrated with polymorphism metrics. 
    
%\end{itemize}

%\begin{minipage}[t]{1\linewidth}
    %\begin{itemize}
\textemdash A variety of structural metrics can represent the internal quality attributes considered in this study. Based on our empirical investigation, for metrics that are associated with quality attributes, there are different degrees of improvement and degradation of software quality.
    %\item Most of the metrics that are mapped to the main quality attributes, \textit{i.e.,} cohesion, coupling, and complexity, do capture developer intentions of quality improvement reported in the commit messages. In contrast, there is also a case in which the metrics do not capture quality improvement as perceived by developers. We summarize our findings as follows:}
    
     \textemdash Most of the metrics that are mapped to the main quality attributes, \textit{i.e.,} cohesion, coupling, and complexity, do capture developer intentions of quality improvement reported in the commit messages. In contrast, there is also a case in which the metrics do not capture quality improvement as perceived by developers. We summarize our findings as follows:
      %\begin{itemize}
        %\item{
        
        \textbf{Cohesion.} In contrast with previous studies, cohesion tends to be well represented by LCOM as we found the metrics to be significantly improved in the refactored code (p-value $\leq$ 0.05).
        %}
        %\item{
        
        \textbf{Coupling.} Similarly to cohesion, optimizing the coupling quality attribute was empirically measured using both CBO and FANOUT (p-value $\leq$ 0.05), in comparison with FANIN and RFC.
        %}
        %\item{
        
        \textbf{Complexity.} One of the popular quality attributes that is being approximated by developers using a variety of metrics, namely CC, WMC, RFC, Evg, NPATH, and MaxNest (p-value $\leq$ 0.05).
        %}
        %\item{
        
        \textbf{Inheritance.} While DIT has been found be the metric that matches the developer's perception (p-value $\leq$ 0.05), NOC, known to be a measure for Inheritance in literature, tends to increase instead in practice (p-value $\leq$ 0.05).
        %}
        %\item{
        
        As for \textbf{Encapsulation}, \textbf{Abstraction} and \textbf{Design Size}. We cannot find any metric that can represent developer's intention of optimizing these quality attributes, and so these findings motivates a deeper investigation on understanding the mismatch between theory and practice.
        %}
      %\end{itemize}
    %\end{itemize}
%\end{minipage}
%
%\begin{minipage}[t]{0.4\linewidth}
%    \begin{itemize}
%      \item{First item}
%      \SubItem{First subitem}
%      \SubItem{Second subitem}
%      \SubItem{Third subitem}
%      \item{Second item}
%      \item{Third item}
%    \end{itemize}
%\end{minipage}
   
As future work, we plan to empirically assess the impact of external quality metrics (\textit{e.g.,} testability and readability) as documented by developers in their commit messages on quality and compare and contrast them with the findings for the internal ones. This will give us an indication which quality attributes are improved the most by developers. Also, we plan on investigating the impact of composed refactorings on each of the quality attributes, in contrast with existing studies which analyze each refactoring type individually. We also want to explore what factors might contribute to the significant improvement of the quality metrics 
%that are aligned with developer perception tagged in the commit messages 
(\textit{e.g.,} developer experience, proximity to release date, and refactoring community culture).

\section*{Acknowledgment}

We sincerely thank the authors of the refactoring mining tools that we have used in this study, for providing their tools open source and for allowing the community to benefit from them.

%\section*{References}

\bibliographystyle{abbrv}
\bibliography{Refactoring}
\begin{comment}

Please number citations consecutively within brackets \cite{b1}. The 
sentence punctuation follows the bracket \cite{b2}. Refer simply to the reference 
number, as in \cite{b3}---do not use ``Ref. \cite{b3}'' or ``reference \cite{b3}'' except at 
the beginning of a sentence: ``Reference \cite{b3} was the first $\ldots$''

Number footnotes separately in superscripts. Place the actual footnote at 
the bottom of the column in which it was cited. Do not put footnotes in the 
abstract or reference list. Use letters for table footnotes.

Unless there are six authors or more give all authors' names; do not use 
``et al.''. Papers that have not been published, even if they have been 
submitted for publication, should be cited as ``unpublished'' \cite{b4}. Papers 
that have been accepted for publication should be cited as ``in press'' \cite{b5}. 
Capitalize only the first word in a paper title, except for proper nouns and 
element symbols.

For papers published in translation journals, please give the English 
citation first, followed by the original foreign-language citation \cite{b6}.

\end{comment}

\end{document}